\documentclass[aps,pra,10pt,twocolumn,superscriptaddress,showpacs]{revtex4}

\usepackage{amsmath}
\usepackage{amssymb}
\usepackage{bbm}
\usepackage{graphicx}
\usepackage{framed}
\usepackage{color}
\usepackage{hyperref}
\usepackage{epstopdf}
\usepackage{dsfont}
\usepackage{amssymb}
\usepackage{amsmath}
\usepackage{amsthm}
\usepackage{wrapfig}
\usepackage{relsize}
\usepackage{bm}
\usepackage[latin1]{inputenc}
\usepackage{enumitem}
\usepackage{natbib}
\usepackage{graphicx}
\usepackage{physics}
\usepackage{hyperref}
\usepackage{xcolor}
\usepackage{subfigure}

\begin{document}

	\title{Optimal Quantum Subtracting machine}
	
	\author{Farzad Kianvash}
	\email{farzad.kianvash@sns.it}
	\author{Marco Fanizza}
	\email{marco.fanizza@sns.it}
	\author{Vittorio Giovannetti}	\affiliation{NEST, Scuola Normale Superiore and Istituto Nanoscienze-CNR, I-56126 Pisa, Italy}
	\date{\today}

	\begin{abstract}
		The impossibility of undoing a mixing process is analysed in the context of quantum information theory. The optimal machine to undo the mixing process is studied in the case of pure states, focusing on qubit systems. Exploiting the symmetry of the problem we parametrise the optimal machine in such a way that the number of parameters grows polynomially in the size of the problem. This simplification makes the numerical methods feasible. 
		For simple but non-trivial cases we computed the analytical solution, comparing the performance of the optimal machine with other protocols.
	\end{abstract}
	
	\maketitle
	\section{Introduction}
	A fundamental fact in quantum information theory is that not all maps between quantum states are possible: even before considering practical difficulties, quantum theory itself limits the operations that can be performed. A series of quantum no-go theorems~\cite{Wootters1982, WERNER,KumarPati2000,PhysRevLett.116.110403,dieks1982communication,PhysRevLett.100.090502, Noprogramming,QA2} shows that transformations which would be very valuable from the point of view of information processing are in fact impossible. The most celebrated of them is the no-cloning theorem~\cite{Wootters1982}: the impossibility of cloning makes many processing tasks (e.g. state estimation) non-trivial. Nonetheless, the importance of these impossible transformations drives the search for approximate implementations of them: optimal cloners ~\cite{RevModPhys.77.1225}  have been extensively studied, and similar efforts have been spent for other no-go theorems~\cite{QA2,PhysRevLett.116.110403,PhysRevA.96.052318,PhysRevA.97.052330, Sepehr}.

	Here we introduce the no-subtracting theorem, which states the impossibility of undoing the mixing operation that involves a target state we wish to recover and an external noise source, and define the optimal subtractor operation which solves the problem with the best allowed approximation.  This task is somehow related to  those discussed in~\cite{ReviewFramesInfo} and references therein, where one aims to 
	perform quantum information processing of some sort (e.g. the recovery of the target state) when some classical knowledge (i.e. the reference frame  for~\cite{ReviewFramesInfo} and
	the amount of mixed noise for us) is replaced  by  bounded information encoded into 
	the density matrix of an ancillary quantum system. Finding the optimal subtractor 
	corresponds to a semidefinite program involving a number of variables that in principle grows 
	exponentially with the input data (system copies).
	However,  by exploiting the symmetry of the problem and a proper parametrisation of the $N$ to 1 qubit covariant channels (analogous to those introduced in Refs.~\cite{ResourceAsymmetry, longevitycovariant}),  the number of effective parameters can be reduced to a subset which only scales polynomially. This reduction of the parameters makes the numerical optimisation feasible, and for small enough input data, allows also for analytical treatment.  
	
As a final consideration we would also to point out that the problem we address in the present paper can also be seen as an instance of quantum machine learning~\cite{QM1,QM2,QM3}, an emerging area of quantum information theory that deals with tasks that generalise the ``learning from example" concept in a genuine quantum information theory setting. In these tasks a machine should be trained to perform a certain quantum operation and this training can be done through quantum processing, that means with quantum training data and quantum operations. The fundamental difference with classical training tasks is that there is not an a priori separation between the training and the execution phases, because of entanglement. Indeed, in our analysis we search for the best machine that can be trained with copies of the noise in order to make it able to recover disturbed signals, with the only requirement that the machine is allowed by quantum mechanics. 
	
	The paper is organized as follows:
we start in Sec.~\ref{SEC2} by formalizing the problem. In Sec.~\ref{SEC3} we present some preliminary results on the efficiency of a universal quantum subtractor which 
can drawn from general consideration on the problem without passing the explicit optimization stage.
In Sec.~\ref{SEC4} we then proceed with the explicit solution of the optimization problem. The paper ends with Sec.~\ref{SEC5}. Technical derivations are presented in the Appendix.

	\section{Optimal Subtractor}  \label{SEC2} 
	An Universal Quantum Subtracting machine UQS is a two-inputs/one-output transformation acting on two isomorphic quantum systems A and B.
	When provided by factorised input states of the form $(p \hat{\rho}_0 + (1-p) \hat{\rho}_1)\otimes \hat{\rho}_0$, with  
	$p\in [0,1]$  assigned  and 
	$\hat{\rho}_0, \hat{\rho}_1\in  \mathfrak{S}{({\cal H})}$ arbitrary density matrices, 
	it  returns as output the system A
	into the state $\hat{\rho}_1$ realizing the mapping 
	\begin{eqnarray} \label{UQS} 
	{\rm UQS}\big[\hat{\rho}_{\rm mix}(p) 
	\otimes \hat{\rho}_0  \big] = \hat{\rho}_1 \label{defQS} \;, 
	\end{eqnarray}  
	which effectively allows one to recover $\hat{\rho}_1$ from the mixture $\hat{\rho}_{\rm mix}(p) := p \hat{\rho}_0 + (1-p) \hat{\rho}_1$ by  ``removing" the perturbing state  $\hat{\rho}_0$  and renormalizing the result.
	Unfortunately  the possibility of physically realizing an UQS machine for $p>0$, 
	turns out to be   in contradiction with the basic requirements that any quantum evolution has to fulfil, see e.g. Ref.~\cite{HOLEVOBOOK}. 
	Indeed invoking  linearity  and using the fact that for  $\hat{\rho}_1=\hat{\rho}_0$ one has ${\rm UQS}[\hat{\rho}_0\otimes \hat{\rho}_0 ] = \hat{\rho}_0$, Eq.~(\ref{UQS}) can be cast in the following form 
	\begin{eqnarray} (1-p){\rm UQS}[\hat{\rho}_1\otimes \hat{\rho}_0 ] =
	{\hat{\rho}_1 - p \hat{\rho}_0}, \end{eqnarray} 
	which, as long as the parameter $p$ is strictly different from 0,
	will produce  unphysical non-positive results as soon as  the support of $\hat{\rho}_0$ admits a non trivial overlap with the kernel of $\hat{\rho}_1$.
	Yet, as in the case of other better studied impossible quantum machines~\cite{WERNER}, there could be still room for 
	approximate implementations of the mapping~(\ref{UQS}). 
	In what follows we shall hence try to  identify the implementation of an optimal UQS,
	i.e. a machine which,  being physically realizable via a Completely Positive and Trace Preserving (CPTP) map~\cite{HOLEVOBOOK}, would give us the best approximation of the transformation~(\ref{UQS}). 
	More generally we are also interested in  a generalisation of the problem where 
	instead of a single copy  of the mixture $p \hat{\rho}_1 + (1-p) \hat{\rho}_0$ and  of the noise state $\hat{\rho}_0$, we are now provided with $n_1$ copies of the first and $n_2$ copies
	of the second, i.e. in the optimal  CPTP implementation of the 
	$\mathfrak{S}{({\cal H}^{\otimes n_1+n_2})}\rightarrow \mathfrak{S}{({\cal H})}$ mapping 
	\begin{eqnarray} \label{UQSn} 
	{\rm UQS}^{(n_1,n_2)}\big[(p \hat{\rho}_0 + (1-p) \hat{\rho}_1)^{\otimes n_1} \otimes \hat{\rho}_0^{\otimes n_2}  \big] = \hat{\rho}_1 \;.
	\end{eqnarray}  
	As a figure of merit we shall consider the fidelity~\cite{DEP1} between the obtained 
	output and the intended target states, properly averaged with respect to all possible inputs.  To simplify the analysis in what follows we restrict ourself to the special case where both $\hat{\rho}_0$ and $\hat{\rho}_1$ are pure states of
	the $d$-dimensional space ${\cal H}$, namely $\hat{\rho}_1= \ket{\psi}\bra{\psi}$ and $\hat{\rho}_0= \ket{\phi}\bra{\phi}$, where  
	without loss of generality  we adopt the parametrisation $\ket{\psi}:=\hat{U} \ket{\uparrow}$ and $\ket{\phi}:=\hat{V} \ket{\uparrow}$, with $\ket{\uparrow}$ being a fixed vector  
	and  $\hat{U}$, $\hat{V}$ are arbitrary elements of the unitary set  $SU(d)$.
	Indicating hence with 
	$\Lambda$ the CPTP mapping
	that we want to test as a candidate for the implementation of ${\rm UQS}^{(n_1,n_2)}$, we evaluate its performance through the function 
	\begin{equation}\label{FIDE}
	F_{n_1,n_2} (\Lambda):= \int \!\!\!\int d {\mu}_U d {\mu}_V
	\bra{\psi}\Lambda[ \hat{\rho}_{\rm mix}^{\otimes n_1}(p)
	\otimes \hat{\rho}_0^{\otimes n_2}
	]\ket{\psi}\;, 
	\end{equation} 
	where the integral  are performed via the Haar measure of $SU(d)$ 
	to ensure a uniform
	distribution of $|\psi\rangle$ and $|\phi\rangle$ on ${\cal H}$. Before entering into the technical derivation, it is worth commenting that while the problem we are facing can be seen as a sort of purification procedure, it is definitely 
	different from the  task  addressed by  Cirac et al. in Ref.~\cite{CIRAC}, which is designed to  remove the largest fraction of complete mixed state from $\hat{\rho}_{{\rm mix}}$ having access to some copies of it, but with no prior information on $\hat{\rho}_0$  or $p$. 
	\subsection{Connection to Quantum Error Correction}
 Equation~(\ref{UQS})  can be described as the formal inversion of  the transformation 
	${\rm IQA}[\hat{\rho}_0\otimes \hat{\rho}_1] =\hat{\rho}_{\rm mix}(p)$, 
	which we may dub Incoherent Quantum Adder.
	At variance with the Coherent Quantum Adder analyzed in Refs.~\cite{QA2,PhysRevA.96.052318,PhysRevA.97.052330}, an IQA can be easily implemented as it merely consists in creating a probability mixture out of two input configurations. 
	In particular, IQA can be interpreted as an open quantum evolution~\cite{HOLEVOBOOK,DEP1,DEP2} in which the state $\hat{\rho}_0$ of the input B, plays the role of the environment. In this scenario, the aim is to undo the action of IQA and recover $\hat{\rho}_1$ not having the full knowledge about the environment, the only information available being  encoded through copies of $\hat{\rho}_0$.
	 In view of this observation  the optimal quantum subtracting problem	can be seen as a first example of a new way of approaching quantum error correction schemes. 
	We are thinking for instance to communication scenarios where, the transformation tampering the received
	state at the output of the channel is affected by interactions with an external environment E that
	is susceptible to modifications on which the communicating parties (say Alice and Bob) do not have a complete record. To be more precise, imagine the following realistic situation where Alice uses a noisy channel to communicate with Bob. Therefore, the states that Alice wants to send to Bob $\rho_1$, interacts with the environment $\rho_0$. The coupling Hamiltonian $H$ between the information carrier $S$ and E and the transfer time $\tau$ of the communication are somehow fixed and known. while the state $\rho_0$ of the environment is not -- $\rho_0$ is a sort of
	a random, possibly time-dependent variable of the problem.
	Accordingly, Bob receives by 
	$\rho_1'=\mbox{Tr}_E[ U_{SE} (\rho_1\otimes \rho_0)U_{SE}^\dag]$ with $U_{SE}= \exp[-i H\tau]$ and $\mbox{Tr}_E[ ... ]$ being the
	partial trace over E. 
	The fundamental task for the receiver of the message is clearly to recover $\rho_1$ from $\rho_1'$: 
	in the standard approach to quantum communication this is facilitated by the assumption that Alice and Bob
	have also perfect knowledge about $\rho_0$ which in our setting  is no longer granted. 
	To compensate for this lack of information it is hence important  for the receiver of the message to  sample the state of E
	in real time during the information exchange, a scenario which we can model e.g. by assuming Bob to
	have access to some copies of $\rho_0$. 
	The optimal subtracting scheme we discuss in the manuscript addresses
	exactly this problem for the special (yet not trivial) case where $U_{SE}$ describes a partial swap gate (see e.g. \cite{scarani3}
	). \\
	Given the above premise it should be now clear that the possibility of constructing an UQS machine will have a profound impact in many practical applications, spanning from
	quantum computation~\cite{QCOMP,DEP1}, where it could be employed as an effective error correction procedure for certain kind of errors,
	to quantum communication~\cite{QCOM}, where instead it could be used as a decoding operation to
	distill the intended messages from the received deteriorated signals.

	\section{Preliminary results} \label{SEC3} The maximum of Eq.~(\ref{FIDE}) with respect to all possible CPTP transformations  
	\begin{eqnarray} \label{MAX}
	{F}_{n_1,n_2}^{(\max)} := \max_{\Lambda\in {\rm CPTP}} F_{n_1,n_2} (\Lambda),\end{eqnarray}  is  the quantity we  are going to study in the following. 
	Since one can always neglect part of the input copies,  
	this functional is clearly non-decreasing in $n_1$ and $n_2$, i.e.  
	\begin{eqnarray} 
	{F}_{n_1,n_2}^{(\max)} \leq {F}_{n_1+1,n_2}^{(\max)},{F}_{n_1,n_2+1}^{(\max)},
	\end{eqnarray}  
	with no 
	ordering between the last two terms been foreseen from first principles.  
	In particular, we are interested in comparing ${F}_{n_1,n_2}^{(\max)}$ with the performances achievable via a trivial ``doing nothing" (DN) strategy in which one emulates the mapping~(\ref{UQSn}) 
	by  simply returning as output one of the qubits of the register A, i.e. the state $\hat{\rho}_{\rm mix}(p)$. 
	In this case, the associated average fidelity  can be easily computed by exploiting the depolarizing identity  
	\begin{eqnarray} \int d {\mu}_U  |\psi\rangle\langle\psi|= \int d {\mu}_U \hat{U} 
	\ket{\uparrow} \bra{\uparrow} \hat{U}^\dag  =\hat{I}/d\;, \end{eqnarray} 
	obtaining  \begin{eqnarray} {F}_{n_1,n_2}({\rm DN}):=1-p (d-1)/d\,, \end{eqnarray} which, by construction constitutes a lower bound for
	${F}_{n_1,n_2}^{(\max)}$, i.e. 
	\begin{eqnarray} {F}_{n_1,n_2}^{(\max)} \geq 1-p (d-1)/d\;,\label{BBBBB}  \end{eqnarray}  (incidentally for the qubit case, ${F}_{n_1,n_2}({\rm DN})$ coincides  with the
	average fidelity on would obtain by adapting the optimal protocol of the Cirac et al. scheme~\cite{CIRAC} to our setting, see Ref.~\cite{SUPMAT}). Determining the exact value of ${F}_{n_1,n_2}^{(\max)}$ is typically very demanding apart from the case where 
	we have a single copy of A, i.e. for $n_1=1$.
	In this scenario in fact, irrespectively from the value of $n_2$,  one can prove that the DN strategy is optimal,
	transforming the inequality~(\ref{BBBBB}) 
	into the identity 
	\begin{eqnarray} 
	{F}_{1,n_2}^{(\max)}   = 1-p (d-1)/d  \label{IDE} \;.
	\end{eqnarray} 
	One way to see this is to show that~(\ref{IDE}) holds in the asymptotic limit of infinitely many copies of the B state, i.e. $n_2\rightarrow 
	\infty$, and then invoke the monotonicity under $n_2$ to extent such result to all the other cases. 
	As a matter of fact when  $n_2$ diverges one can use quantum  tomography to  recover the 
	classical description of B from the input data: accordingly the optimal implementation of ${\rm UQS}^{(1,\infty)}$ formally coincides with the optimal recovery map~\cite{IPPO}  aiming to invert
	the CPTP transformation that takes a generic element $\hat{\rho}_1 \in \mathfrak{S}{({\cal H})}$ into  $\hat{\rho}_{\rm mix}(p)$.
	In this case~(\ref{FIDE}) gets replaced by  
	\begin{equation} F_{1,\infty} (\Lambda):= \int  d {\mu}_U  \bra{\psi}\Lambda[(1-p)\ket{\psi}\bra{\psi}
	+p\ket{\phi}\bra{\phi}
	]\ket{\psi}\;,\end{equation} 
	which thanks to the depolarizing identity can be easily shown to admit 
	${F}_{n_1,n_2}({\rm DN})$
	not just as a lower bound but also as an upper bound, leading to 
	\begin{eqnarray} {F}_{1,\infty}^{(\max)}   = 1-p (d-1)/d\;, \end{eqnarray} and hence to~(\ref{IDE}).

	As $n_1$ gets larger than $1$, we aspect to see a non trivial improvement with respect to the DN strategy. 
	This is clearly evident at least in the case where  both $n_1$ and $n_2$ diverge (i.e. $n_1,n_2\rightarrow \infty$).
	In this regime, similarly to the case of optimal quantum cloner~\cite{Wootters1982,OPTIMAL1,OPTIMAL2,OPTIMAL3,OPTIMAL4,OPTIMAL5}, 
	Eq.~(\ref{UQSn}) becomes implementable by means of
	a simple measure-and-prepare (MP) strategy based on 
	performing full quantum tomography on both inputs A and B, yielding the optimal value $F_{\infty,\infty}^{(\max)}=1$ which clearly
	surpasses the DN threshold.
	In the next sections, we shall clarify a procedure that one can follow to solve the optimisation  of  
	Eq.~(\ref{FIDE}) for finite values of the input copies. For the sake of simplicity we present it for the special cases where A and B are just qubit systems and we use such technique to analytically compute the exact value of 
	${F}_{n_1,n_2}^{(\max)}$ for the simplest but non-trivial scenario where $n_1=2$ and $n_2=1$. Via numerical methods we also solve the optimisation problem for some selected values of 
	$p$, $n_1$ and $n_2$, see Fig.~\ref{colormaps}.
	\section{Channel optimisation} \label{SEC4} 
	The problem we are considering has  special symmetries that allows for some simplifications. 
	Invoking the linearity of $\Lambda$ and the invariance of the Haar measure we can 
	rewrite  (\ref{FIDE}) as
	$F_{n_1,n_2}(\Lambda)=\bra{\uparrow}{\Lambda_c}[\hat{\Omega}_{n_1n_2}]\ket{\uparrow}$,
	where $\hat{\Omega}_{n_1n_2}$ is the density operator 
	\begin{multline} \label{DEFOMEGA} 
	\hat{\Omega}_{n_1n_2}:=\int d {\mu}_V \left(p\ket{\uparrow}\bra{\uparrow}+(1-p)\hat V \ket{\uparrow}\bra{\uparrow} \hat V ^\dagger \right)^{\otimes n_1}
	\\\otimes \left(\hat V \ket{\uparrow}\bra{\uparrow} \hat V ^\dagger\right)^{\otimes n_2}.
	\end{multline}
	The channel  ${\Lambda_c}$ appearing in the expression for $F_{n_1,n_2}(\Lambda)$
	is obtained from $\Lambda$ through the following  integral 
	\begin{equation} \label{FORM} 
	{\Lambda_c}[\cdots]=\int d{\mu}_U  \; \hat{U}\Lambda[\hat{U}^{\dagger^{\otimes N}} \cdots \hat{U}^{\otimes N}]\hat{U}^\dagger\;,
	\end{equation}
	which ensures that  ${\Lambda_c}$ is a  $N$ qubits to 1 qubit covariant map, i.e. a  CPTP transformation 
	fulfilling the condition
	\begin{equation} \hat{U}^\dag{\Lambda_c}[\cdots] \hat{U}= {\Lambda_c}[\hat{U}^{\dagger^{\otimes N}}\cdots \hat{U}^{\otimes N}]\;,\end{equation}  $\forall
	\hat{U} \in SU(2)$~\cite{COVA}.
	Notice also that if $\Lambda$ is already covariant, then it coincides with its associated ${\Lambda_c}$, i.e.
	${\Lambda_c}=\Lambda$.
	Exploiting these facts we can hence conclude that 
	the maximisation of  
	$F_{n_1,n_2}(\Lambda)$ can be performed  by just focusing on this special set of transformations 
	which now we shall 
	parametrise.
	The  integral  appearing in~(\ref{FORM}) 
	motivates us to choose the total angular momentum eigenbasis as the basis for the  Hilbert space ${\cal H}_2^{\otimes N}$ where 
	the channel operates. Specifically we shall write  such vectors as 
	$\ket{{j,m,g}}$
	with $j$ the total angular momentum of $N$ spin $1/2$  particles, $m$ the total angular momentum in $z$ direction, and $g$ labelling different equivalent representations with total angular momentum $j$. 
	Following the derivation presented in~\cite{SUPMAT}  we can then verify that, indicating with 
	$\{\ket{j=\tfrac{1}{2}, s}\}_{s =\pm \tfrac{1}{2}}$ the  angular momentum  basis for a single qubit (no degeneracy being present), 
	one has 
	\begin{eqnarray}  \label{charcterisation} &&
	\bra{{\tfrac{1}{2},s}} 
	{\Lambda_c}\left[
	\big|j,m,g\rangle\langle j^{\prime},m^{\prime},g^{\prime}\big|
	\right] \big|{\tfrac{1}{2}}, {s}'\big\rangle=
	(-1)^{m-m'}  \\ 
	&&    \nonumber
	\quad  \times  \delta_{{s}-m,{s}'-m'} \;  \sum_{q \in Q_{j,j'}} 
	C_{\frac{1}{2} \, {s},j\, -m}^{q \quad {s} -m}\, C_{\frac{1}{2} \, {s}',j'\,  -m'}^{q \quad {s}'-m'}
	\;\;  W^{j,j'}_{q,g,g'} \;, 
	\end{eqnarray}
	where  the summation over the index $q$ runs over  
	\begin{eqnarray} Q_{j,j'}:=\{j\pm \frac{1}{2}\}\cap \{j'\pm \frac{1}{2}\}\;, \end{eqnarray} (if the set is empty then the associated matrix element is automatically null), where
	\begin{eqnarray} C^{J\quad M}_{j\, m, j'\, m'}:=\langle {J,M}|j, m\rangle\otimes | j' , m'\rangle\;, \end{eqnarray} 
	is a Clebsch-Gordan coefficient, and where finally \begin{eqnarray} W^{j,j'}_{q,g,g'} :=
	\mathbf{v}^{j}_{q,g} \cdot (\mathbf{v}^{j'}_{q,g'})^\dag\;,\end{eqnarray} 
	represents the scalar product between the complex row vectors 
	$\mathbf{v}^{j}_{q,g}$ and $\mathbf{v}^{j'}_{q,g'}$ 
	constructed from the Kraus operators of $\Lambda$ and explicitly defined in~\cite{SUPMAT}. Equation~(\ref{charcterisation}) tells us which are the parameters characterizing  ${\Lambda_c}$ that enter into the optimization problem. The number of $W^{j,j'}_{q,g,g'}$ grows exponentially in $n_1$ and $n_2$: the multiplicity of the representation with total angular momentum $j$ grows exponentially in general, therefore $g$ and $g'$ can take an exponential number of different values.  
	It is worth observing that  this quantity 
	does not depend on  $m,s,m',s'$ which only appear in the Clebsch-Gordan coefficients. 
	Also, the structure of the covariant channels specified in Eq.~(\ref{charcterisation}) indicates that the action of ${\Lambda_c}$ on the off-diagonal elements in the total angular momentum basis is zero unless $\abs{m-m'}=1$ and $\abs{j-j'}=1$. In principle there is no selection rule on $g$ and $g'$, and at this level the number of variables of the problem still scales exponentially in $n_1$ and $n_2$. A dramatic simplification arises by using the symmetry properties of $\hat{\Omega}_{n_1n_2}$. First of all $[\hat{\Omega}_{n_1n_2},\hat J_z]=0$, from which $\bra{j,m,g}\hat{\Omega}_{n_1n_2}\ket{j',m',g'}$ is zero unless $m=m'$. Moreover, by Schur-Weyl duality \cite{FultonHarris} the Hilbert space of the problem can be decomposed as
	$\mathcal H= \bigoplus_{D_1,D_2} (j_{D_1} \otimes \alpha_{D_1}) \otimes (j_{D_2}\otimes \alpha_{D_2})$, 
	where $j_{D_i}$ and $\alpha_{D_i}$ are the irreducible representations of $SU(2)$ and the symmetric group $S_{n_i}$ with Young diagram $D$. We notice that $\hat{\Omega}_{n_1n_2}$ is symmetric under permutations acting independently on the first $n_1$ and the second $n_2$ qubits, hence by Schur's lemma $\hat{\Omega}_{n_1n_2}$ must have the form
	$\hat{\Omega}_{n_1n_2}= \bigoplus_{D_1,D_2} (\hat\Omega_{D_1} \otimes \mathbf 1_{D_1}) \otimes (\hat\Omega_{D_2}\otimes  \mathbf 1_{D_2})$,
	with $\hat\Omega_{D_1},\hat\Omega_{D_2}$ positive-semidefinite operators.
	In particular,
	since $(\hat V \ket{\uparrow}\bra{\uparrow}\hat V^\dagger)^{\otimes n_2}$ is supported on the completely symmetric subspace for each $\hat V$, 
	$\hat\Omega_{D_2}=0$ unless $D_2$ is the completely symmetric Young diagram.
	From this observation it follows that $\hat{\Omega}_{n_1n_2}$ is supported on a space spanned by orthonormal vectors labelled as $\ket{j,m, g_{j_{1}}}$, where we use the same conventions as before for the total angular momentum indices, and we simplify the notation using $g_{j_1}$ as a shortcut for the couple $(j_{D_1},g_{D_1})$ which indexes a basis of $\alpha_{D_1}$. Putting all together we have proved that
	$\langle{j,m, {g_{j_{1}}}}|\hat{\Omega}_{n_1n_2}|{j',m', g'_{j'_{1}}}\rangle={\Omega}_{n_1n_2}(j,j',m,j_1,p)\delta_{m,m'}\delta_{g_{j_1},g'_{j'_1}}$,
	with the function ${\Omega}_{n_1n_2}(j,j',m,j_1,p)$ depending only on $j,j',m,j_1,p$ and being
	explicitly computed in \cite{SUPMAT}.
	Exploiting these properties of $\hat{\Omega}_{n_1n_2}$ the fidelity can then be expressed as
	\begin{equation}\label{Fidelityn1n2} 
	F_{n_1,n_2}(\Lambda)=\sum_{j,j',{j_1},q}C^{j,j'}_{q,{j_1}}(p)W^{j,j'}_{q,{j_1}} \, ,
	\end{equation}
	where $C^{j,j'}_{q,{j_1}}(p)$ is a contraction of Clebsch-Gordan coefficients defined in~\cite{SUPMAT}, and 
	\begin{eqnarray} W^{j,j'}_{q,{j_1}}:=({\sum_{g_{j_1}}W^{j,j'}_{q, g_{j_1}, g_{j_1}}})/{\#g_{j_1}}\;,\end{eqnarray} with \begin{eqnarray} 
	\#g_{j_1}=\frac{(n_1)!(2{j_1}+1)}{(\frac{n_1-2{j_1}}{2})(\frac{n_1+2{j_1}}{2}+1)}\;,\end{eqnarray}  being the multiplicity of the representation $j_1$. 
	Further constraints associated with the Completely Positivity condition of  ${\Lambda_c}$ are also automatically included
	in the  parametrisation via $W^{j,j'}_{q,g,g'}$ 
	through the connection between  the vectors 
	$\mathbf{v}^{j}_{q,g}$ and the Kraus operators of 
	$\Lambda$. The trace preserving requirement reduces instead to 
	$\sum_{{s}= \pm \tfrac{1}{2}}
	\langle\tfrac{1}{2},{s}|
	{\Lambda_c}\left[|{j,m,g}\rangle\langle{j',m',g'}|\right]
	|\tfrac{1}{2},{s}\rangle=\delta_{j,j'}\delta_{m,m'}\delta_{g,g'}$,
	which via some manipulations~\cite{SUPMAT} can be cast in the equivalent form 
	\begin{eqnarray}\label{constraint2}
	\tfrac{2+2j}{(1+2j)} \; W^{j,j}_{q,j_1} \Big\rvert_{{q=j+\tfrac{1}{2}}} 
	+\tfrac{2j}{1+2j}\; 
	W^{j,j}_{q,j_1} \Big\rvert_{q=j-\tfrac{1}{2}}
	=  1\;.
	\end{eqnarray}
	We notice that 
	the linearity of $F_{n_1,n_2}(\Lambda)$ and the convexity of the set of channels allows us to
	restrict the search for the maximum fidelity among those $\Lambda$s for which $W^{j,j'}_{q, g_{j_1}, g'_{j_1}}=\delta_{g_{j_1}, g'_{j_1}}W^{j,j'}_{q,j_1}$. Accordingly, the latter become the effective variables over which one  has to
	perform the 
	maximization of~(\ref{Fidelityn1n2}). As explicitly shown in~\cite{SUPMAT} their number grows polynomially in $n_1$ and $n_2$, reducing the problem  to 
	a semidefinite program which let us perform numerical optimisation.
	\subsection{Results} 
	\begin{figure}[t!]
		\includegraphics[width=1\columnwidth]{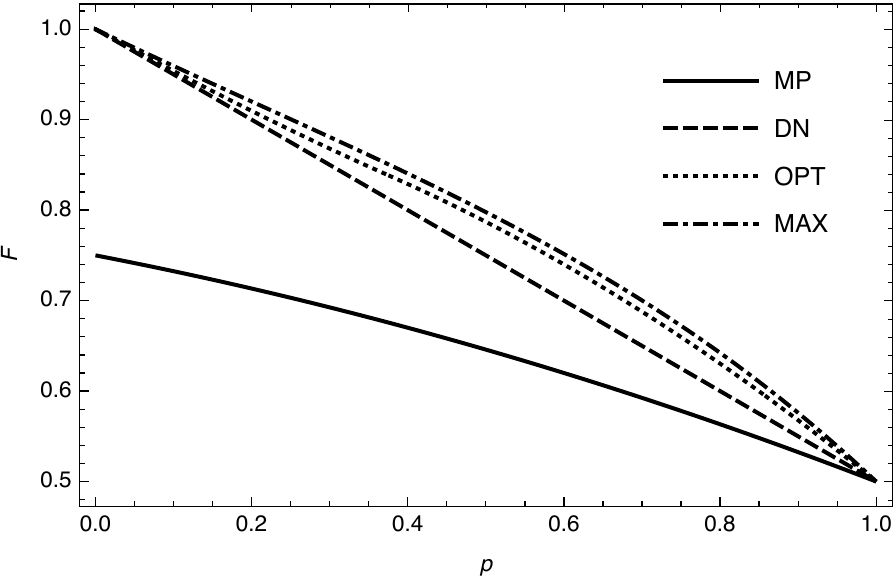}
		\caption{Average fidelity~(\ref{OPTIF}) for the optimal UQS machine for $n_1=2$ and $n_2=1$  as a function of the probability 
			parameter $p$ (dotted line), together with the average fidelities of the DN strategy (dashed line),  the optimal expression for $n_1=2$ and
			$n_2=\infty$ corresponding to have classical knowledge of the perturbing state (dotted-dashed line), and the upper bound $F_{n_1=2}^{\rm upper}(\Lambda^{\text{MP}})$
			attainable with  MP procedures (continuous line).
			\label{figure2new}}
	\end{figure}
\begin{figure}[t!]
		\begin{tabular}{ c c }
			\includegraphics[width=0.5\columnwidth]{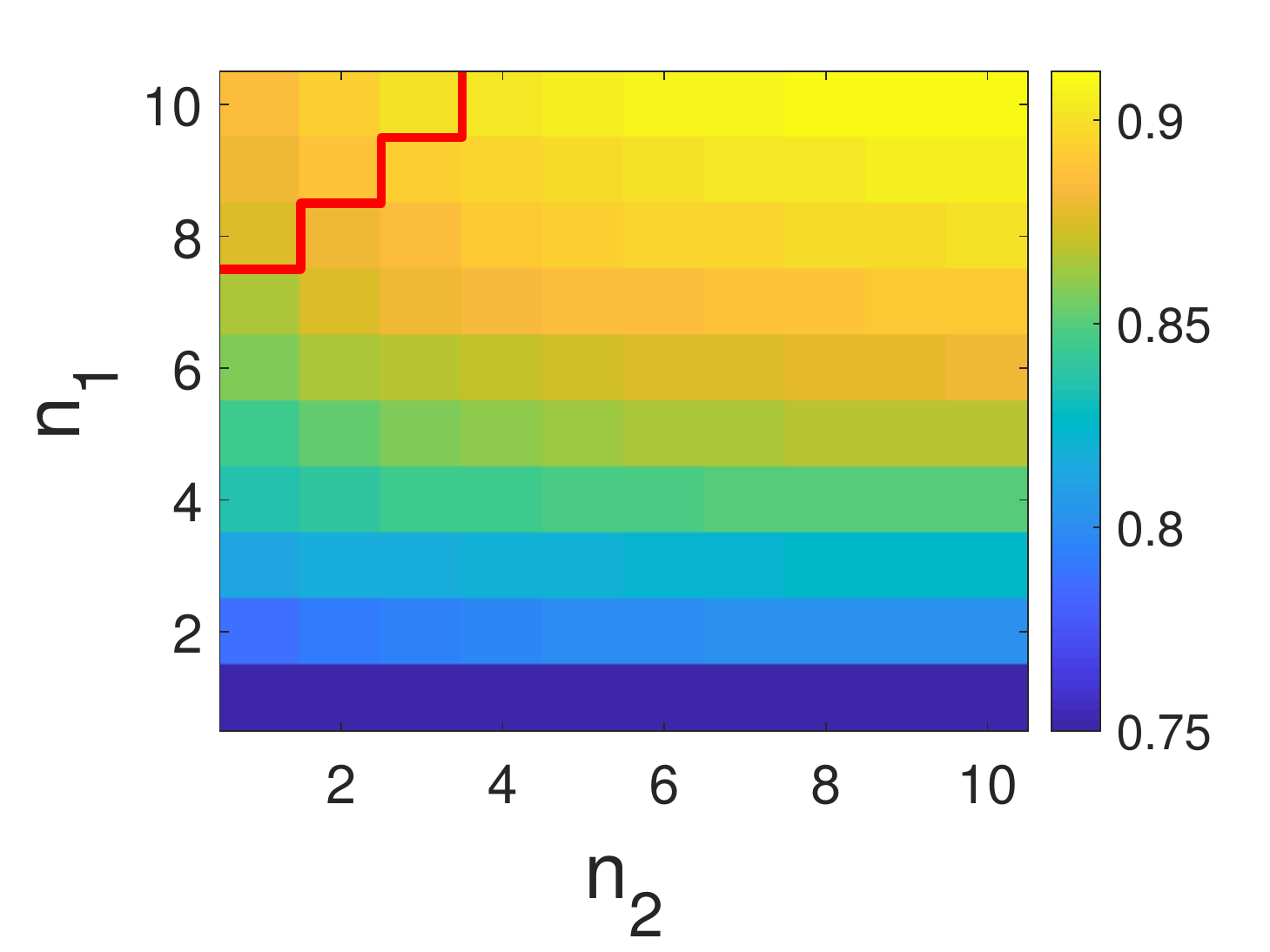} & \includegraphics[width=0.5\columnwidth]{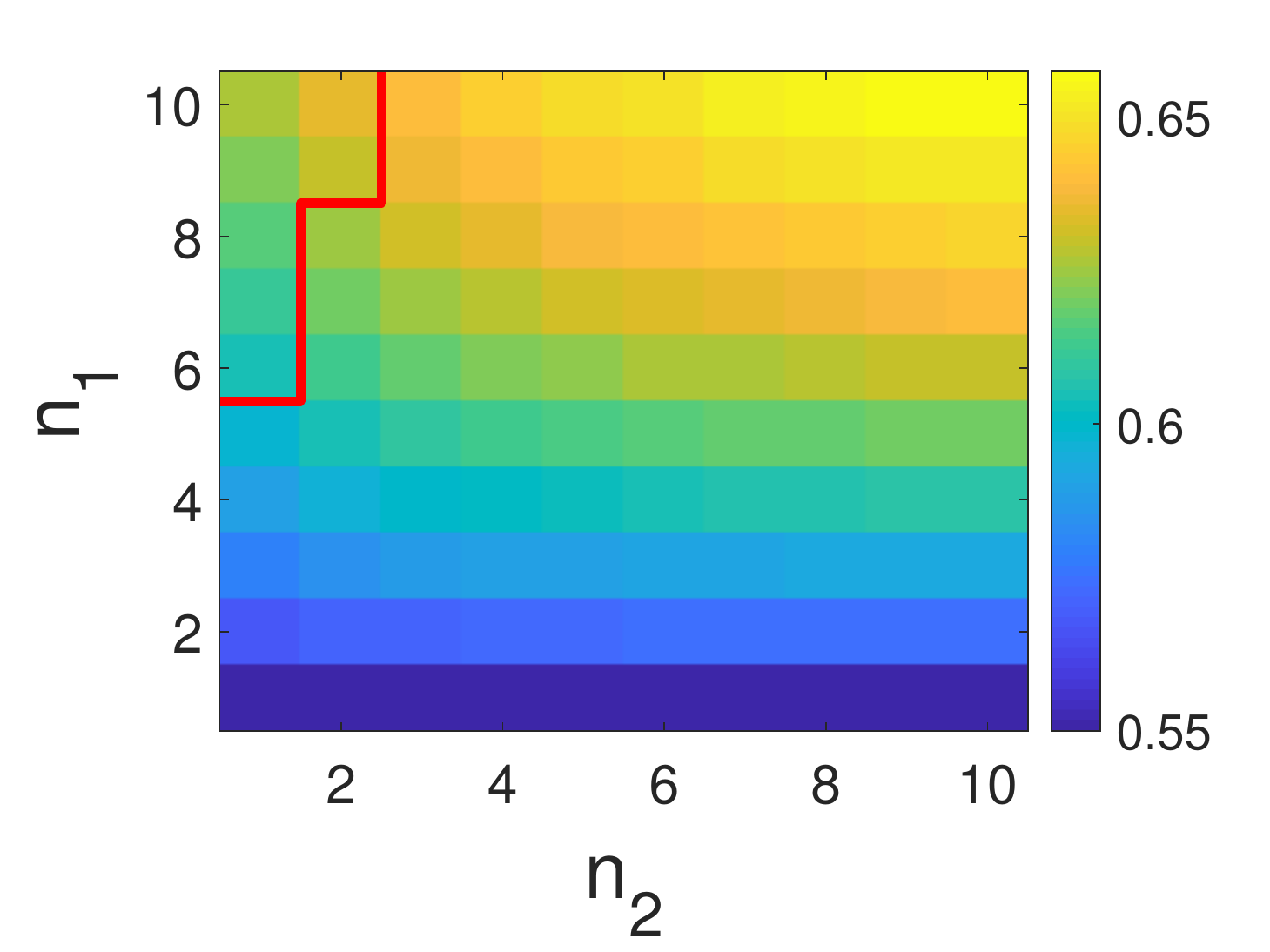}
		\end{tabular}
		\caption{Numerical results for the maximum  fidelity for $p=0.5$ (left) and $p=0.9$ (right) for $n_1=1,..,10$ and $n_2=1,..,10$. The red line indicates the following: as long as $(n_2,n_1+1)$ is below the red line, ${F}_{n_1+1,n_2}^{(\max)}>{F}_{n_1,n_2+1}^{(\max)}$, otherwise  ${F}_{n_1+1,n_2}^{(\max)}<{F}_{n_1,n_2+1}^{(\max)}$. For example, in $p=0.5$ (left), for $n_1=8$ and $n_2=2$, the fidelity is larger if we add one copy of the input B instead of the input A.
			\label{colormaps}}
	\end{figure}

	As explicitly shown in~\cite{SUPMAT} the maximization of Eq.~(\ref{Fidelityn1n2})  for case $n_1=2$ and $n_2=1$ can be performed analytically leading to 
	\begin{equation}
	F_{2,1}^{(\max)}=
	\begin{cases}
	\frac{(1-p)(51+23p)}{54}+\frac{(1-p)(3+p)^2}{27(6-7p)}+\tfrac{p^2}{2},&  0\leq p\leq \frac{3}{8}\\
	\frac{(1-p)(51+23p)}{54}+\tfrac{p(1-p)}{3}+\frac{p^2}{2},           &  \frac{3}{8}\leq p \leq 1\;.
	\end{cases} \label{OPTIF} 
	\end{equation}
	In Fig.~\ref{figure2new} we report Eq.~(\ref{OPTIF}) together with the average fidelity for the DN strategy, 
	with the function $F_{2,\infty}^{(\max)}$  which we computed in~\cite{SUPMAT} 
	following the same approach used for  ${F}_{1,\infty}^{(\max)}$,
	and with 
	the curve
	\begin{eqnarray} F_{n_1=2}^{\rm upper}(\Lambda^{\text{MP}}):= (9-2p-p^2)/12\;, \end{eqnarray}
	which, as we detail in~\cite{SUPMAT}, provides an upper bound to the average fidelity attainable 
	when resorting on MP strategies  when having $n_1=2$ copies of A and arbitrary copies of $\hat{\rho}_0$. The curves show
	that for the low value $n_1$  we are considering here, 
	the MP procedures are ineffective even with respect to the trivial DN strategy.  
	$F_{2,1}^{(\max)}$ on the contrary is strictly larger than the DN score. Also it is very close to $F_{2,\infty}^{(\max)}$, showing that for $n_1=2$, the possibility of having just a single copy of the perturbing
	state $\hat{\rho}_0$  provides us almost all the benefit one could obtain by having a classical knowledge of the latter.
	For larger values of $n_1$  and $n_2$ analytical treatment becomes cumbersome and we resort to
	numerical analysis using  Mathematica \cite{Mathematica} 
	to compute the parameters of the problem and  CVX, a package for specifying and solving convex programs \cite{MATLAB, CVX1,CVX2} in Matlab, to calculate the maximum fidelity values. 
	Results are reported in Fig.~\ref{colormaps} 
	for $n_1=1,2,..,10$ and $n_2=1,2,..,10$ and $p=\frac{1}{2},\frac{9}{10}$. 
	
	\section{Conclusions}\label{SEC5} 
	The  gap between $F_{2,1}^{(\max)}$ and the DN strategy (which is optimal in the $n_1=1$ scenario and independent from the explicit value of  $n_2$)  shows that even a small redundancies on the input A, 
	can be beneficial.
	On the contrary, the very small distance between 
	$F_{2,1}^{(\max)}$ and $F_{2,\infty}^{(\max)}$ clarifies that gathering more information on the mixing term $\hat{\rho}_0$ (the noise of the model) does 
	not help too much. 
	As can be seen from Fig.~\ref{colormaps} for larger $n_1$ and $n_2$ one can instead see a noise-dependent separation line between two regions, one where it is indeed advantageous to increase $n_1$ instead of $n_2$ and the other where the opposite holds.
	
	The symmetry of the problem allowed us to reduce exponentially the number of variables involved in the optimisation. The same analysis should be relevant also in a broader perspective for general noise models.
	\\
	\section{Acknowledgments} The first two authors contributed equally to this work.

\newpage

\begin{widetext}
\section{Appendix}
Supplemental material is organised as following. First, we provide explicit derivation of the decomposition~(\ref{charcterisation}) of the main text. Then using ~(\ref{charcterisation}) we derive the fidelity~(\ref{Fidelityn1n2}) of the main text. Third, we present an explicit derivation of Eq.~(\ref{constraint2}) of the main text. Then
	we analyze the application of the decomposition~(\ref{charcterisation}) of the main text to the case where $n_1=n_2=1$, and in the following section we do the same for the case $n_1=2, n_2=1$. 
	Analytical optimisation is also done for the case $n_1=2, =\infty$. Then, we present the derivation of an upper bound for the average fidelity of the UQS realised via 
	measurement and prepare strategies. Finally, we apply the method of Cirac et al.~\cite{CIRAC} for case $n_1=2$ and arbitrary $n_2$. 
\subsection{Covariant Channel Characterisation} \label{App B}
Here the calculations to derive the characterisation for covariant are presented.
Introducing a Kraus decomposition for $\Lambda$ in Eq.(\ref{FORM}) of the main text  we get
\begin{eqnarray}\label{Kraus}
	{\Lambda_c}[\cdots]=\sum_k \int d{\mu}_U \; \hat{U} \hat{M}_k \hat{U}^{\dagger^{\otimes N}}
	[\cdots] \hat{U}^{\otimes N}\hat{M}_{k}^{\dagger}\hat{U}^{\dagger}\;,
\end{eqnarray}
with $\hat{M}_k$  the associated Kraus operators. Accordingly we can express the 
matrix element~(\ref{charcterisation})  
\begin{equation} \label{characterisation 2}
\bra{\tfrac{1}{2},{s}}{\Lambda_c}(\ket{j,m,g}\bra{j',m',g'})\ket{\tfrac{1}{2},{{s}'}}=  \sum_{k}\sum_{r,r'} \sum_{l,l'}  \int d{\mu}_U
  D^{1/2}_{{s},r}(\hat{U}) M^{k,j,g}_{r,l}D^{j}_{l,m}(\hat{U}^{\dagger})D^{j'}_{m',l'}(\hat{U}) M^{\dagger^{k,j',g'}}_{l',r'}D^{1/2}_{r',{{s}'}}(\hat{U}^\dagger)\;, 
\end{equation}
where
\begin{equation}
D^j_{l,l'}(\hat{U}):=\bra{j,l,g}\hat{U}^{\otimes N}\ket{j,l',g}\;, \qquad 
M^{k,j,g}_{r,l}:=\bra{\tfrac{1}{2},r}\hat{M}_k\ket{j,l,g}\;. 
\end{equation}
 We can write the multiplication of two Wigner matrices in the following form
\begin{eqnarray}
D^{1/2}_{{s},r}(\hat{U})D^{j}_{l,m}(\hat{U}^{\dagger})&=&(-1)^{l-m}\bra{\tfrac{1}{2},{s}}\hat{U}\ket{\tfrac{1}{2},r}\bra{j,-m,g}\hat{U}^{\otimes N}\ket{j,-l,g} \nonumber \\
&=&(-1)^{l-m}\bra{1/2,{s}}\otimes\bra{j,-m,g}\hat{U}^{\otimes N+1}\ket{1/2,r}\otimes\ket{j,-l,g}\nonumber\\
&=&  (-1)^{l-m} \sum_{j-\frac{1}{2}\leq q\leq j+\frac{1}{2} } C_{\frac{1}{2} \, {s},j\, -m}^{q \quad {s} -m}\, C_{\frac{1}{2} \, r,j\, -l}^{q \quad r-l} \,D^q_{{s}-m,r-l}(\hat{U}) \;, 
\end{eqnarray}
where $C^{J\quad M}_{j\, m, j'\, m'}=\bra{J,M}\ket{j\, m}\otimes \ket{ j' \, m'}$ are the Clebsch-Gordan coefficients. 
Exploiting this we can hence rewrite Eq.~(\ref{characterisation 2}) in the following form
\begin{eqnarray} \label{eq2}
\bra{\tfrac{1}{2},{s}}{\Lambda_c}(\ket{j,m,g}\bra{j',m',g'})\ket{\tfrac{1}{2},{{s}'}} 
&=& \sum_k \sum_{r,r'} \sum_{l,l'} \sum_{q,q'}  (-1)^{l-m+l'-m'}    \int d\mu_U \quad C_{\frac{1}{2} \, {s},j\, -m}^{q \quad {s} -m}\, C_{\frac{1}{2} \, r,j\, -l}^{q \quad r-l} \nonumber \\
&&\times 
D^q_{{s}-m,r-l}(\hat{U}) M^{k,j,g}_{r,l}C_{\frac{1}{2} \, r',j'\, -l'}^{q' \quad r'-l'}\, C_{\frac{1}{2} \, {{s}'},j'\, -m'}^{q' \quad {{s}'}-m'} 
D^{q'}_{r'-l',{{s}'}-m'}(\hat{U}^\dagger)M^{\dagger^{k,j',g'}}_{l',r'}\;.
\end{eqnarray}
Remembering that  following identity of Wigner matrices (Peter-Weyl theorem, see \cite{Knapp})
\begin{equation}
\int
d\mu_U\, {D^j_{m,l}(\hat{U})D^{j'}_{m',l'}}(\hat{U})^*=\frac{1}{2j+1}\delta_{j,j'}\delta_{m,m'}\delta_{l,l'}
\end{equation}
the integral in~(\ref{eq2}) can hence be simplified to
\begin{eqnarray} \label{eq3}
\bra{\tfrac{1}{2},{s}}{\Lambda_c}(\ket{j,m,g}\bra{j',m',g'})\ket{\tfrac{1}{2},{{s}'}}&=&  \sum_k \sum_{r,r'} \sum_{l,l'} \sum_{q} \frac{(-1)^{l-m+l'-m'}}{2j+1}  
\delta_{{s}-m,{{s}'}-m'}\delta_{r-l,r'-l'} \nonumber \\
&&\times C_{\frac{1}{2} \, {s},j\, -m}^{q \quad {s} -m}\, C_{\frac{1}{2} \, {{s}'},j'\, -m'}^{q \quad {{s}'}-m'} \, C_{\frac{1}{2} \, r,j\, -l}^{q \quad r-l} 
M^{k,j,g}_{r,l}C_{\frac{1}{2} \, r',j'\, -l'}^{q \quad r'-l'}\, M^{\dagger^{k,j',g'}}_{l',r'} \;.
\end{eqnarray}
Introducing then the variable $p:=r-l=r'-l'$, we can rewrite the above identity as 
\begin{eqnarray}\label{characterisation 3}
\bra{1/2,{s}}{\Lambda_c}(\ket{j,m,g}\bra{j',m',g'})\ket{1/2,{{s}'}}&=& \sum_k \sum_{r,r'} \sum_{p} \sum_{q} \frac{ (-1)^{r-m+r'-m'}}{2j+1}  \delta_{{s}-m,{{s}'}-m'} 
\nonumber \\
&&C_{\frac{1}{2} \, {s},j\, -m}^{q \quad {s} -m}\, C_{\frac{1}{2} \, {{s}'},j'\, -m'}^{q \quad {{s}'}-m'}  C_{\frac{1}{2} \, r,j\, p-r}^{q \quad p} \, M^{k,j,g}_{r,r-p}C_{\frac{1}{2} \, r',j'\, p-r'}^{q \quad p}\, M^{\dagger^{k,j',g'}}_{r'-p,r'}\;,
\end{eqnarray}
which, defining the row vectors ${\bf v}^j_{q,g}$ of components 
\begin{equation}\label{def_f}
{\bf v}^j_{q,g}({k,p}):=\frac{1}{2 j+1} \sum_r {(-1)^r} C_{\frac{1}{2} \, r,j\, p-r}^{q \quad p} \, M^{k,j,g}_{r,r-p}\;, 
\end{equation}
and their associated scalar products 
\begin{eqnarray}\label{defw} W^{j,j'}_{q,g,g'} :=
  \mathbf{v}^{j}_{q,g} \cdot (\mathbf{v}^{j'}_{q,g'})^\dag
  = \sum_{k,p}{\bf v}^j_{q,g}({k,p})
\left[{\bf v}^{j'}_{q',g'}({k,p})\right]^*\;, 
  \end{eqnarray} 
allows us to finally express Eq. (\ref{characterisation 3}) as in Eq.~(\ref{charcterisation}) of the main text
\begin{eqnarray}  
\bra{{\tfrac{1}{2},s}} 
{\Lambda_c}\left[
\big|j,m,g\rangle\langle j^{\prime},m^{\prime},g^{\prime}\big|
\right] \big|{\tfrac{1}{2}}, {s}'\big\rangle=
(-1)^{m-m'}   \delta_{{s}-m,{s}'-m'} \;  \sum_{q \in Q_{j,j'}} 
C_{\frac{1}{2} \, {s},j\, -m}^{q \quad {s} -m}\, C_{\frac{1}{2} \, {s}',j'\,  -m'}^{q \quad {s}'-m'}
\;\;  W^{j,j'}_{q,g,g'} \;.
\end{eqnarray} 
  \subsection{Fidelity calculation for arbitrary $n_1,n_2$ and numerical optimisation}
 Using Eq.~(\ref{DEFOMEGA}) of the main text the average fidelity can be expressed as 
\begin{equation} 
F_{n_1,n_2}(\Lambda_c)=\bra{\uparrow}{\Lambda_c}[\int d {\mu}_V \left(p\ket{\uparrow}\bra{\uparrow}+(1-p)\hat V \ket{\uparrow}\bra{\uparrow} \hat V ^\dagger \right)^{\otimes n_1}
\\\otimes \left(\hat V \ket{\uparrow}\bra{\uparrow} \hat V ^\dagger\right)^{\otimes n_2}]\ket{\uparrow} \;,
\end{equation}
 Knowing that $\Omega_{n_1,n_2}$ is invariant under any permutation on the first $n_1$ qubits, we can write
\begin{equation} 
F_{n_1,n_2}(\Lambda_c)=\frac{1}{\abs{S_{n_1}}}\bra{\uparrow}{\Lambda_c}[\sum_{\sigma}\hat{\Pi}_{\sigma} \int d {\mu}_V \left(p\ket{\uparrow}\bra{\uparrow}+(1-p)\hat V \ket{\uparrow}\bra{\uparrow} \hat V ^\dagger \right)^{\otimes n_1}
\\\otimes \left(\hat V \ket{\uparrow}\bra{\uparrow} \hat V ^\dagger\right)^{\otimes n_2}\hat{\Pi}^\dagger_{\sigma}]\ket{\uparrow} \;,
\end{equation}
where $\hat{\Pi}_{\sigma}$ is a permutation on the first $n_1$ qubits, and $\sigma$ runs over all the elements of the symmetric group $S_{n_1}$, and $\abs{S_{n_1}}$ is the number of elements of symmetric group. Then we can write
\begin{equation} \label{Fidelity 4}
F_{n_1,n_2}(\Lambda_c)=\frac{1}{\abs{S_{n_1}}}\bra{\uparrow}{\Lambda_c}[\sum_{k=0}^{n_1}\sum_{\sigma} {n_1 \choose k}  (1-p)^k p^{n_1-k} \hat{\Pi}_{\sigma} \ket{\uparrow}\bra{\uparrow}^{\otimes k}\otimes \hat{A}_{N-k}\hat{\Pi}^\dagger_{\sigma}]]\ket{\uparrow} \;,
\end{equation}
where $\hat{A}_{k}:=\int d {\mu}_V [\hat{V} \ket{\uparrow}\bra{\uparrow}\hat{V}^\dag]^{\otimes k}$. Defining $\hat{B_k}$, we carry on the calculation
\begin{eqnarray}
\hat{B_k}&:=&\ket{\uparrow}\bra{\uparrow}^{\otimes k}\otimes \hat{A}_{N-k}\\ \nonumber
&=&\sum_{m,m',s,s'}\frac{\delta_{m+s,m'+s'}}{N-k+1}C_{\frac{N-k-n_2}{2}\,{m}, \, {\frac{n_2}{2}}\, s}^{{\frac{N-k}{2}} \quad m+s}C_{\frac{N-k-n_2}{2}\,{m'}, \, {\frac{n_2}{2}}\, s'}^{{\frac{N-k}{2}} \quad m'+s'}  \ket{\uparrow}\bra{\uparrow}^{\otimes k}\\ \nonumber
&&\otimes\ket{\frac{N-k-n_2}{2},m}\bra{\frac{N-k-n_2}{2},m'}\otimes\ket{\frac{n_2}{2}, s}\bra{\frac{n_2}{2}, s'}\, .
\end{eqnarray}
Note that here we do not need to sum over any multiplicity index for the states $\ket{\frac{N-k-n_2}{2},m}$ and $\ket{\frac{n_2}{2}, s}$, because $\hat{A}_{N-k}$ is supported on the completely symmetric subspace of $N-k$ qubits, therefore it is also supported on the tensor product of the completely symmetric subspaces of $N-k-n_2$ and $n_2$ qubits, which have multiplicity 1. Writing the first $n_1$ qubits in the total angular momentum basis we get
\begin{eqnarray}
\hat{B_k}&=& \sum_{m,s,m',s',{j_1},{j_1}'}\frac{\delta_{m+s,m'+s'}}{N-k+1}C_{\frac{N-k-n_2}{2}\,{m}, \, {\frac{n_2}{2}}\, s}^{\frac{N-k}{2} \quad m+s}C_{\frac{N-k-n_2}{2}\,{m'}, \, {\frac{n_2}{2}}\, s'}^{{\frac{N-k}{2}} \quad m'+s'} C_{\frac{k}{2}{\frac{k}{2}}, \, {\frac{N-k-n_2}{2}}\, m}^{{j_1} \quad \frac{k}{2} +m} C_{\frac{k}{2}{\frac{k}{2}}, \, {\frac{N-k-n_2}{2}}\, m'}^{{j_1}' \quad \frac{k}{2} +m'}\\ \nonumber
&&\ket{{j_1},\frac{k}{2}+m,k}\bra{{j_1}',\frac{k}{2}+m',k}\otimes\ket{\frac{n_2}{2}, s}\bra{\frac{n_2}{2}, s'}\, 
\end{eqnarray}
here the multiplicity index $k$ indicates that we first wrote the $k$ qubits in the total angular momentum basis then we summed it up with $\ket{\frac{N-k-n_2}{2},m}\bra{\frac{N-k-n_2}{2},m}$. Schur's lemma implies
\begin{equation}\label{shur's lemma}
\frac{1}{\abs{S_{n_1}}}\sum_{\sigma}\hat{\Pi}_{\sigma}\ket{{j_1},m,k}\bra{{j_1}',m',k}\hat{\Pi}^\dagger_{\sigma}=\sum_{g_{j_1}}\frac{1}{\# g_{j_1}}\ket{{j_1},m,g_{j_1}}\bra{{j_1},m',g_{j_1}} \delta_{{j_1},{j_1}'} \, ,
\end{equation}
where $g_{j_1}$ is the index for the multiplicity of ${j_1}$ and runs over all the possible values for a certain $j_1$, and $\#g_{j_1}=\frac{(n_1)!(2{j_1}+1)}{(\frac{n_1-2{j_1}}{2})(\frac{n_1+2{j_1}}{2}+1)}$. Using Eq.~(\ref{shur's lemma}) in Eq.~(\ref{Fidelity 4}) we get
\begin{eqnarray}
F_{n_1,n_2}(\Lambda_c)
&=&\bra{\uparrow}{\Lambda_c}\bigg[ \sum_{m,s,m',s',{j_1},k,g_{j_1}}\frac{1}{\#g_{j_1}}{n_1 \choose k}  (1-p)^k p^{n_1-k} \frac{\delta_{m+s,m'+s'}}{N-k+1}C_{\frac{N-k-n_2}{2}\,{m}, \, {\frac{n_2}{2}}\, s}^{{\frac{N-k}{2}} \quad m+s}C_{\frac{N-k-n_2}{2}\,{m'}, \, {\frac{n_2}{2}}\, s'}^{{\frac{N-k}{2}} \quad m'+s'}\\ \nonumber
&& C_{\frac{k}{2}{\frac{k}{2}}, \, {\frac{N-k-n_2}{2}\, m}}^{{j_1} \quad \frac{k}{2} +m}C_{\frac{k}{2}{\frac{k}{2}}, \, {\frac{N-k-n_2}{2}\, m'}}^{{j_1} \quad \frac{k}{2} +m'}\ket{{j_1},\frac{k}{2}+m,g_{j_1}}\bra{{j_1},\frac{k}{2}+m',g_{j_1}}\otimes\ket{\frac{n_2}{2}, s}\bra{\frac{n_2}{2}, s'}\bigg]\ket{\uparrow}\, \\ \nonumber
&=&\bra{\uparrow}{\Lambda_c}\bigg[ \sum_{m,s,m',s',{j_1},j,j',k,g_{j_1}}\frac{1}{\#g_{j_1}}{n_1 \choose k}  (1-p)^k p^{n_1-k} \frac{\delta_{m+s,m'+s'}}{N-k+1}C_{\frac{N-k-n_2}{2}\,{m}, \, {\frac{n_2}{2}}\, s}^{{\frac{N-k}{2}} \quad m+s}C_{\frac{N-k-n_2}{2}\,{m'}, \, {\frac{n_2}{2}}\, s'}^{{\frac{N-k}{2}} \quad m'+s'}\\ \nonumber
&& C_{\frac{k}{2}{\frac{k}{2}}, \, {\frac{N-k-n_2}{2}\, m}}^{{j_1} \quad \frac{k}{2} +m}C_{\frac{k}{2}{\frac{k}{2}}, \, {\frac{N-k-n_2}{2}\, m'}}^{{j_1} \quad \frac{k}{2} +m'} C_{j_1{\frac{k}{2}+m}, \, {\frac{n_2}{2}}\, s}^{{j} \quad \frac{k}{2} +m+s}C_{j_1{\frac{k}{2}+m'}, \, {\frac{n_2}{2}}\, s'}^{{j'} \quad \frac{k}{2} +m'+s'}\ket{{j},\frac{k}{2}+m+s,g_{j_1}}\bra{{j'},\frac{k}{2}+m'+s',g_{j_1}}\bigg]\ket{\uparrow}\, .
\end{eqnarray}
Using the Eq.~(\ref{charcterisation}) of the main text we get
\begin{eqnarray} \label{Fidelity8}
F_{n_1,n_2}(\Lambda_c)&=& \sum_{m,s,m',s',{j_1},j,j',k,g_{j_1},q}\frac{1}{\#g_{j_1}}{n_1 \choose k}  (1-p)^k p^{n_1-k}\frac{\delta_{m+s,m'+s'}}{N-k+1}C_{\frac{N-k-n_2}{2}\,{m}, \, {\frac{n_2}{2}}\, s}^{{\frac{N-k}{2}} \quad m+s}C_{\frac{N-k-n_2}{2}\,{m'}, \, {\frac{n_2}{2}}\, s'}^{{\frac{N-k}{2}} \quad m'+s'}\\ \nonumber
&& C_{\frac{k}{2}{\frac{k}{2}}, \, {\frac{N-k-n_2}{2}\, m}}^{{j_1} \quad \frac{k}{2} +m}C_{\frac{k}{2}{\frac{k}{2}}, \, {\frac{N-k-n_2}{2}\, m'}}^{{j_1} \quad \frac{k}{2} +m'} C_{j_1{\frac{k}{2}+m}, \, {\frac{n_2}{2}}\, s}^{{j} \quad \frac{k}{2} +m+s}C_{j_1{\frac{k}{2}+m'}, \, {\frac{n_2}{2}}\, s'}^{{j'} \quad \frac{k}{2} +m'+s'}C_{\frac{1}{2}{\frac{1}{2}}, \, {j\, -\frac{k}{2}-m-s}}^{{q} \quad -\frac{k-1}{2}-m-s}C_{\frac{1}{2}{\frac{1}{2}}, \, {j'\, -\frac{k}{2}-m'-s'}}^{{q} \quad -\frac{k-1}{2}-m'-s'}W^{j,j'}_{q,g_{j_1},g_{j_1}}\, .
\end{eqnarray}
The dependence of the coefficients of $W^{j,j'}_{q,g_{j_1},g_{j_1}}$ on the multiplicity index $g_{j_1}$ is only through $j_1$. So, we can define
\begin{eqnarray}
C^{j,j'}_{q,{j_1}}(p):=&&\sum_{m,s,m',s',k}{n_1 \choose k}  (1-p)^k p^{n_1-k}\frac{\delta_{m+s,m'+s'}}{N-k+1}C_{\frac{N-k-n_2}{2}\,{m}, \, {\frac{n_2}{2}}\, s}^{{\frac{N-k}{2}} \quad m+s}C_{\frac{N-k-n_2}{2}\,{m'}, \, {\frac{n_2}{2}}\, s'}^{{\frac{N-k}{2}} \quad m'+s'}\\ \nonumber
&& C_{\frac{k}{2}{\frac{k}{2}}, \, {\frac{N-k-n_2}{2}\, m}}^{{j_1} \quad \frac{k}{2} +m}C_{\frac{k}{2}{\frac{k}{2}}, \, {\frac{N-k-n_2}{2}\, m'}}^{{j_1} \quad \frac{k}{2} +m'} C_{j_1{\frac{k}{2}+m}, \, {\frac{n_2}{2}}\, s}^{{j} \quad \frac{k}{2} +m+s}C_{j_1{\frac{k}{2}+m'}, \, {\frac{n_2}{2}}\, s'}^{{j'} \quad \frac{k}{2} +m'+s'}C_{\frac{1}{2}{\frac{1}{2}}, \, {j\, -\frac{k}{2}-m-s}}^{{q} \quad -\frac{k-1}{2}-m-s}C_{\frac{1}{2}{\frac{1}{2}}, \, {j'\, -\frac{k}{2}-m'-s'}}^{{q} \quad -\frac{k-1}{2}-m'-s'}\, ,
\end{eqnarray}
and write the fidelity as 
\begin{eqnarray}\label{Fidelity7}
F_{n_1,n_2}(\Lambda_c)=\sum_{j,j',{{j_1}},q}C^{j,j'}_{q,{j_1}}(p)\sum_{g_{j_1}} \frac{1}{\#g_{j_1}} W^{j,j'}_{q,g_{j_1},g_{j_1}} \, ,
\end{eqnarray}
Because $\Omega_{n_1,n_2}$ is symmetric on the first $n_1$ qubits, we can always choose $\Lambda_{c}$ to be symmetric on the first $n_1$ qubits, therefore
\begin{eqnarray}
	\Lambda_{c}[\hat{\rho}]=\frac{1}{\abs{S_{n_1}}}\sum_{\sigma}\Lambda_{c}[\hat{\Pi}_{\sigma}\hat{\rho}\hat{\Pi}^\dagger_{\sigma}]\, .
\end{eqnarray}
Using~(\ref{shur's lemma}) we derive 
\begin{eqnarray}
	W^{j,j'}_{q,g_{j_1},g_{j_1}}=\frac{1}{\#g_{j_1}}\sum_{g'_{j_1}}W^{j,j'}_{q,g'_{j_1},g'_{j_1}}\, ,
\end{eqnarray}
therefore
\begin{equation}\label{sym equality}
	W^{j,j'}_{q,g_{j_1},g_{j_1}}=W^{j,j'}_{q,g'_{j_1},g'_{j_1}} \quad \forall g_{j_1}, g_{j'_1}\, .
\end{equation}
  so defining $W^{j,j'}_{q,{j_1}}:=\frac{1}{\#g_{j_1}}\sum_{g_{j_1}}W^{j,j'}_{q,g_{j_1},g_{j_1}}$, then we get
\begin{eqnarray}\label{Fidelity9}
F_{n_1,n_2}(\Lambda_c)=\sum_{j,j',{j_1},q}C^{j,j'}_{q,{j_1}}(p)W^{j,j'}_{q,{j_1}} \, .
\end{eqnarray}

Now, the number of parameters i.e. $W^{j,j'}_{q,{j_1}}$, scale polynomially with $n_1$ and $n_2$ because the multiplicity index is fixed to be $j_1$ and the number of different $j_1$ is $\mathcal O(n_1)$. Without using the characterisation of covariant channels and writing $\Omega_{n_1,n_2}$ in the proper form, the number of parameters grows exponentially in $n_1$ and $n_2$. This exponential reduction of parameters makes the numerical optimisation feasible. In fact, this optimisation problem is exactly a semidefinite programming optimisation. To show this we first briefly review the semidefinite programming and then we define the parameters in the program.\\
A general semidefinite program can be defined as any mathematical program of the form \cite{SDP}
\begin{eqnarray}
&&\max_{\hat{X}\in \mathbb{S}^n} \quad  F_{n_1,n_2}(\hat{X})=\Tr[\hat{C}^\intercal \hat{X}]\\ \nonumber
&&\text{subject to} \quad \Tr[\hat{D}^\intercal_k \hat{X}]\geq b_k, \quad k=1,..,m, \quad \text{and} \quad \hat{X}\geq 0
\end{eqnarray}
where $\mathbb{S}^n$ is the space of all real $n\times n$ matrices. $\hat{C}$ and $\hat{D_k}$ are $n\times n$ real matrices, and $b_k$ are real numbers and $\hat{X}\geq 0$ means that $\hat{X}$ is semidefinite.\\
In our problem, $C^{j,j'}_{q,{j_1}}(p)$ are the matrix elements of $\hat{C}$ which are all real since $C^{j,j'}_{q,{j_1}}(p)$ is the combination of the Clebsch-Gordan coefficients. Our constraints are equality constraints, each of which can be obtained from two inequalities. The matrix elements of $\hat{X}$ are $W^{j,j'}_{q,{j_1}}$, and the elements of $\hat{D_k}$ and $b_k$ can be read from the coefficients in Eq.~(\ref{constraint2}) of the main text.\\
To prove that our problem is a semidifinite program we should show that $\hat{X}$ is positive-semidefinite. $\hat{X}$ is positive-semidefinite if and only if there exists a set of vectors like $\{ v_i \}$ such that $x_{m,n}=v^\intercal_m . v_n$. In the definition of $W^{j,j'}_{q,{g,g'}}$ in Eq.~(\ref{defw}), we have 
\begin{eqnarray} W^{j,j'}_{q,{g_{j_1}},g_{j_1}} :=
\mathbf{v}^{j}_{q,g_{j_1}} \cdot (\mathbf{v}^{j'}_{q,g_{j_1}})^\dag\, , 
\end{eqnarray}
and using Eq.~(\ref{sym equality}) we get 
\begin{eqnarray}
 W^{j,j'}_{q,{{j_1}},{j_1}} :=
\mathbf{v}^{j}_{q,{j_1}} \cdot (\mathbf{v}^{j'}_{q,{j_1}})^\dag\, . 
\end{eqnarray}
So, $\hat{X}\geq0$ and our problem is a semidefinite program.\\
Note that in our maximisation problem the parameters in general can be complex numbers. However, the matrix elements of $\hat{C}$ are the contraction of Clebsch-Gordan coefficients which are all real, therefore without loss of generality we can assume that $W^{j,j'}_{q,{j_1}}$ are real.
\subsection{Derivation of Eq.~(17)}
Here we give explicit derivation of the constraint~(15) of the main text. 
The starting point to observe that by explicit substitution of Eq.~(13) into Eq.~(14)  of the main text 
we get 
\begin{equation}
	\sum_{{s}= \pm \tfrac{1}{2}} \sum_q  C_{\frac{1}{2} \, {s},j\, -m}^{q \quad {s} -m}\, C_{\frac{1}{2} \, {s},j'\,  -m}^{q \quad {s}-m} \;\;
	 W^{j,j'}_{q,g,g'} 
	=\delta_{j,j'}\delta_{g,g'}\;. \label{EQ1} 
\end{equation}
Using then the following symmetry property of Clebsch-Gordan coefficients
\begin{equation}
C_{j_1 \, {m_1},j_2\, m_2}^{J M}=(-1)^{j_1-m_1}\sqrt{\tfrac{2J+1}{2j_2+1}}C_{j_1 \, {m_1},J\, -M}^{j_2 -m_2}\;,
\end{equation}
we can observe that 
\begin{eqnarray}
	\sum_{{s}= \pm \tfrac{1}{2}} C_{\frac{1}{2} \, {s},j\, -m}^{q \quad {s} -m}\, C_{\frac{1}{2} \, {s},j'\,  -m}^{q \quad {s}-m} =\sum_{{s}= \pm \tfrac{1}{2}}\frac{2q+1}{2j+1}C_{\frac{1}{2} \, {s},q\, m-s}^{j \quad {m}}C_{\frac{1}{2} \, {s},q\, m-s}^{j '\quad {m} }=\frac{2q+1}{2j+1}\bra{j,m}\hat{\hat{\Pi}}_m\ket{j'm}=\frac{2q+1}{2j+1}\delta_{j,j'},
\end{eqnarray}
where $\hat{\hat{\Pi}}_m$ is the projector on the the $j_z=m$ eigenspace.
It follows hence that~(\ref{EQ1}) is automatically fulfilled for 
 $j\ne j'$, while for $j=j'$ instead it gives (17) of the main text
 \begin{eqnarray}\label{constraint3}
 \tfrac{2+2j}{1+2j} \; W^{j,j}_{q,g,g'} \Big\rvert_{{q=j+\tfrac{1}{2}}} 
 +\tfrac{2j}{1+2j}\; 
 W^{j,j}_{q,g,g'} \Big\rvert_{q=j-\tfrac{1}{2}}
 =  \delta_{g,g'}\;.
 \end{eqnarray}
 Using the definition of $W^{j,j'}_{q,{j_1}}:=\frac{1}{\#g_{j_1}}\sum_{g_{j_1}}W^{j,j'}_{q,g_{j_1},g_{j_1}}$ and summing the equations~(\ref{constraint3}) we get
 \begin{eqnarray}
 \tfrac{2+2j}{(1+2j)} \; W^{j,j}_{q,j_1} \Big\rvert_{{q=j+\tfrac{1}{2}}} 
 +\tfrac{2j}{1+2j}\; 
 W^{j,j}_{q,j_1} \Big\rvert_{q=j-\tfrac{1}{2}}
 =  1\;.
 \end{eqnarray}

\subsection{Application of the formalism to the case $n_1=n_2=1$}   \label{App C}
For $n_1=n_2=1$, Eq.~(\ref{DEFOMEGA}) of the main text explicitly yields
	\begin{eqnarray}
	\hat{\Omega}_{1,1}=(1-p)\ket{\uparrow}\bra{\uparrow}\otimes {\hat{I}}/{2}+p \hat{A}_2. \end{eqnarray} 
	Notice that the  term 
 $\hat{A}_2$ is invariant under rotations hence it gets mapped by $\Lambda_c$ into a multiple of the identity operator: specifically noticing that  $\mbox{Tr}[\hat{A}_2]=1$ we have 
 $\Lambda_{c}[\hat{A}_2] = \int d{\mu}_U  \; \hat{U}\Lambda[\hat{A}_2]\hat{U}^\dagger = \hat{I}/2$ which implies 
 $\bra{\uparrow}{\Lambda_c}\big[\hat{A}_2\big]\ket{\uparrow} =1/2$.
 On the contrary the first contribution to $\hat{\Omega}_{1,1}$ 
 admits the following decomposition 
 \begin{equation}
 	\ket{\uparrow}\bra{\uparrow}\otimes {\hat{I}}=|{1,1}\rangle\langle{1,1}|+\frac{|{1,0}\rangle+|{0,0}\rangle}{\sqrt{2}}
	\frac{\langle{1,0}|+\langle{0,0}|}{\sqrt{2}}\;,
 \end{equation}
 where  
without loss of generality we identified $\ket{\uparrow}$ with the vector $|\tfrac{1}{2},\tfrac{1}{2}\rangle$,  
and 
 where in the r.h.s. appear states of the total angular momentum basis of two spin $\frac{1}{2}$ (no multiplicity being present). Using  Eq.~(\ref{charcterisation}) of the main text and the table of Clebsch-Gordan coefficients we can then write
  \begin{eqnarray}
	\bra{\uparrow}{\Lambda_c}\big[\ket{\uparrow}\bra{\uparrow}\otimes {\hat{I}}\big]\ket{\uparrow}&=&\frac{2}{3}\abs{
	\mathbf{v}^{1}_{{3}/{2}}}^2+\frac{5}{6}\abs{\mathbf{v}^{1}_{{1}/{2}}}^2  
	+\frac{1}{2}\abs{\mathbf{v}^{0}_{{1}/{2}}}^2
	\nonumber \\ &&+\frac{1}{\sqrt{3}}\Re{\mathbf{v}^{0}_{{1}/{2}}\cdot (\mathbf{v}^{1}_{{1}/{2}})^\dag}\;, 
 \end{eqnarray}
 where we dropped the index $g$ since here is no multiplicity in total angular momentum basis of two qubits. 
 Similarly the constraints~(\ref{constraint2}) of the main text  becomes
    \begin{equation}
	 \frac{4}{3}\abs{	\mathbf{v}^{1}_{{3}/{2}}}^2+\frac{2}{3}\abs{\mathbf{v}^{1}_{{1}/{2}}}^2 =1, \quad 
	 2\abs{\mathbf{v}^{0}_{{1}/{2}}}^2=1\;.
	 \label{CONST} 
 \end{equation}
Exploiting this we observe that  fidelity of $F_{1,1}(\Lambda)$ for a generic map  must fulfil  the constraint 
  \begin{eqnarray} \label{fidelity 3}
 F_{1,1}(\Lambda)&\leq &\frac{p}{2}+\frac{1-p}{2}\bigg[\frac{2}{3}\abs{
	\mathbf{v}^{1}_{{3}/{2}}}^2+\frac{5}{6}\abs{\mathbf{v}^{1}_{{1}/{2}}}^2  
	+\frac{1}{2}\abs{\mathbf{v}^{0}_{{1}/{2}}}^2
	\nonumber \\ &&+\frac{1}{\sqrt{3}} \abs{\mathbf{v}^{0}_{{1}/{2}}}\abs{\mathbf{v}^{1}_{{1}/{2}}}\bigg] \leq 1- p/2\;,
 \end{eqnarray}
 the first inequality being obtained by forcing  $\mathbf{v}^{0}_{{1}/{2}}$ and $\mathbf{v}^{1}_{{1}/{2}}$ to be collinear, while the second following directly from (\ref{CONST}).
 By comparing this with the lower bound ${F}_{n_1,n_2}^{(\max)} \geq 1-p (d-1)/d$ discussed in the main text 
 for the qubit case (i.e. $d=2$) this allows us to recover the identity~(\ref{IDE}) of the main text, i.e. 
   \begin{eqnarray} 
{F}_{1,1}^{(\max)}   = 1-p /2  \label{IDEqubit} \;,
\end{eqnarray} 
the bound being achived by employing the DN strategy. 
\subsection{Details of the Calculation for $n_1=2,n_2=1$}   \label{App D}
 Here we present detailed calculation to derive Eq.~(18) of the main text. Using the Eq.~(\ref{Fidelity8}) we can write the fidelity as 
\begin{eqnarray} 
\nonumber 
F_{2,1}(\Lambda)&=&\tfrac{p^2}{2} + \tfrac{5(1-p)(3+5p)
	 {W^{{3}/{2},{3}/{2}}_{2,1}}
 +(1-p)(33+23p)W^{{3}/{2},{3}/{2}}_{1,1}}{72}+ \tfrac{p(1-p)W^{{1}/{2},{1}/{2}}_{0,0}+5 p(1-p)W^{{1}/{2},{1}/{2}}_{1,0}}{12}+\tfrac{(1-p)(3+p) W^{{3}/{2},{1}/{2}}_{1,1}}{9\sqrt{2}}  \nonumber \\ 
&&+\tfrac{(6-p)(1-p)W^{{1}/{2},{1}/{2}}_{1,1}+(1-p)(6-5p)W^{{1}/{2},{1}/{2}}_{0,1}}{36}\;,
\end{eqnarray}
using the definition of $W^{j,j'}_{q,{j_1}}:=\frac{1}{\#g_{j_1}}\sum_{g_{j_1}}W^{j,j'}_{q,g_{j_1},g_{j_1}}$ and the definition of $W^{j,j'}_{q,g,g'}$ in Eq.~(\ref{defw}) we can write
\begin{eqnarray} 
\nonumber 
F_{2,1}(\Lambda)&=&\tfrac{p^2}{2} + \tfrac{5(1-p)(3+5p) \abs{\mathbf{v}^{{3}/{2}}_{2,1}}^2+(1-p)(33+23p)\abs{\mathbf{v}^{{3}/{2}}_{1,1}}^2}{72}+ \tfrac{p(1-p)\abs{
		\mathbf{v}^{{1}/{2}}_{0,2}}^2+5 p(1-p)\abs{\mathbf{v}^{{1}/{2}}_{1,2}}^2}{12}+\tfrac{(1-p)(3+p)\abs{
		\mathbf{v}^{{3}/{2}}_{1,1}
	}\abs{
		\mathbf{v}^{{1}/{2}}_{1,1}}}{9\sqrt{2}}  \nonumber \\ 
&&+\tfrac{(6-p)(1-p)\abs{\mathbf{v}^{{1}/{2}}_{1,1}}^2+(1-p)(6-5p)\abs{\mathbf{v}^{{1}/{2}}_{0,1}}^2}{36}\;,
\label{NEWEQ1} 
\end{eqnarray}
with constraints:
\begin{eqnarray} 
\tfrac{{5 \abs{\mathbf{v}^{3/2}_{2,1}}^2+3\abs{\mathbf{v}^{3/2}_{1,1}}^2}}{4}=1\;, \qquad 
\tfrac{{3}\abs{\mathbf{v}^{1/2}_{1,g}}^2+\abs{\mathbf{v}^{1/2}_{0,g}}^2}{2} =1\;, \label{CONSTR1} 
\end{eqnarray}
where $g=1,2$.
Using the constraints we eliminate $\abs{\mathbf{v}^{3/2}_{2,1}},\abs{\mathbf{v}^{1/2}_{0,2}},\abs{\mathbf{v}^{1/2}_{0,1}}$, Eq.~(\ref{NEWEQ1}) becomes
\begin{eqnarray} 
F_{2,1}(\Lambda)=\tfrac{3-2p(1-p)}{6} + \tfrac{p(1-p)\abs{\mathbf{v}^{{1}/{2}}_{1,2}}^2}{6}+\tfrac{(1-p)(3+p)\abs{\mathbf{v}^{{3}/{2}}_{1,1}}^2}{9}+\tfrac{(1-p)(3+p)\abs{
		\mathbf{v}^{{3}/{2}}_{1,1} \label{FFDAM}
	}\abs{\mathbf{v}^{{1}/{2}}_{1,1}}}{9\sqrt{2}}-\tfrac{(1-p)(6-7p)\abs{\mathbf{v}^{{1}/{2}}_{1,1}}^2}{18}  \;.
\end{eqnarray}
The coefficients of  $\abs{\mathbf{v}^{1/2}_{1,2}},\abs{\mathbf{v}^{3/2}_{1,1}}$ are positive everywhere, so to maximise the fidelity we put their maximum values $\abs{\mathbf{v}^{{1}/{2}}_{1,2}}^2=\frac{2}{3},\abs{\mathbf{v}^{3/2}_{1,1}}^2=\frac{4}{3}$, obtaining
\begin{eqnarray} 
F_{2,1}(\Lambda)=\tfrac{51-4p(7-p)}{54} +\tfrac{\sqrt{2}(1-p)(3+p)\abs{\mathbf{v}^{{1}/{2}}_{1,1}}}{9\sqrt{3}}-\tfrac{(1-p)(6-7p)\abs{\mathbf{v}^{{1}/{2}}_{1,1}}^2}{18}  \;. 
\end{eqnarray}
This last expression has to be maximise with respect to $\abs{\mathbf{v}^{{1}/{2}}_{1,1}}$ considering that, according to
the constraint~(\ref{CONSTR1}) such variable has to belong to the interval $[0,\sqrt{2/3}]$. 
 For the case $p<\frac{6}{7}$ we take the derivative and put it equal to zero obtaining 
\begin{eqnarray} \label{vmax}
\abs{\mathbf{v}^{{1}/{2}}_{1,1}}=\tfrac{\sqrt{2}(3+p)}{\sqrt{3}(6-7p)}  \;,
\end{eqnarray}
which belongs to the allowed interval only when  $p<{3}/{8}$. Accordingly for these values of
$p$ we can use Eq.~(\ref{vmax}) obtaining  
\begin{eqnarray} \label{OPTIFappe0} 
F^{(\max)}_{2,1}=\tfrac{51-4p(7-p)}{54}+\tfrac{(1-p)(3+p)^2}{27(6-7p)} \;.
\end{eqnarray}
For the case $1>p>{3}/{8}$ (which incidentally also includes $1>p>{6}/{7}$), instead the maximum 
for (\ref{FFDAM}) is always maximised 
for the maximum allowed value of  $\abs{\mathbf{v}^{{1}/{2}}_{1,1}}$, i.e. $\abs{\mathbf{v}^{{1}/{2}}_{1,1}}=\sqrt{2/3}$ yielding
\begin{equation}
	F_{2,1}^{(\max)}=
	\frac{(1-p)(51+23p)}{54}+\tfrac{p(1-p)}{3}+\frac{p^2}{2},      
 \label{OPTIFappe} 
\end{equation}
which together with (\ref{OPTIFappe0}) gives us (18) of the main text.

\subsection{Case $n_1=2$, $n_2=\infty$} \label{App E}

As we argued in the text, 
\begin{equation}\label{MAX_F_INFINITY}
{F}_{n_1,\infty}^{(\max)}= \max_{\Lambda\in {\rm CPTP}}  \int d {\mu}_U
	\bra{\psi}\Lambda[\hat{\rho}_{\rm mix}^{\otimes n_1}(p)]\ket{\psi}=\max_{\Lambda\in {\rm CPTP}}  \int d {\mu}_U
	\bra{\psi}\Lambda[ \left((1-p)\ket{\psi}\bra{\psi}+p\ket{\phi}\bra{\phi}\right)^{\otimes 2}]\ket{\psi}\;, 
\end{equation}
This equality is consistent since ${F}_{n_1,\infty}^{(\max)}$ does not depend on $\ket{\phi}$ by virtue of the invariance property of the Haar measure, and therefore one can set $\ket{\phi}=\ket{0}$ without loss of generality. In this case the optimal $\Lambda$ is not covariant, since it depends on $\ket{0}\bra{0}$, but we can still find the maximum fidelity through the standard Kraus representation of $\Lambda$. For $n_2=2$, there is no need to distinguish between equivalent representations and the matrix elements $M^{(k)}_{s,j,m}$ of a set of Kraus operators for $\Lambda$, $\hat M_k$, satisfy
\begin{align}
\bra{\tfrac 1 2 , s}\Lambda[\ket{j,m}\bra{j',m'}]\ket{\tfrac 1 2, s'}&=\sum_{k}M^{(k)}_{s,j,m}\overline{{M}^{(k)}_{s',j',m'}}\\
\sum_{s=-1/2}^{1/2}\bra{\tfrac 1 2 , s}\Lambda[\ket{j,m}\bra{j',m'}]\ket{\tfrac 1 2, s}&=\sum_{s=-1/2}^{1/2}\sum_{k}M^{(k)}_{s,j,m}\overline{{M}^{(k)}_{s,j',m'}}=\delta_{j,j'}\delta_{m,m'}.
\end{align}

The integral in (\ref{MAX_F_INFINITY}) can be written as
\begin{align}
&\int d {\mu}_U
	\bra{\psi}\Lambda[ \left((1-p)\ket{\psi}\bra{\psi}+p\ket{0}\bra{0}\right)^{\otimes 2}]\ket{\psi}=\nonumber\\
	&=\sum_{s,s',l,m,l',m'}\int d {\mu}_U\bra{\tfrac 1 2 , s}D_{\tfrac 12,s}^{\tfrac 1 2}(\hat{U}^\dagger)\Lambda[{\hat{\rho}_{\rm mix}^{\otimes 2}(p)}_{l,m,l',m'}\ket{\tfrac 12, l}\bra{\tfrac 12, m}\otimes\ket{\tfrac 12, l'}\bra{\tfrac 12, m'}]D_{s',\tfrac 12}^{\tfrac 1 2}(\hat{U})\ket{\tfrac 1 2, s'},\nonumber\\
	&=\sum_{s,s',l,m,l',m'}\int d {\mu}_U{\hat{\rho}_{\rm mix}^{\otimes 2}(p)}_{l,m,l',m'}D_{\tfrac 12,s}^{\tfrac 1 2}(U^\dagger)D_{s',\tfrac 12}^{\tfrac 1 2}(\hat{U})\left(\sum_{k,j,j'} M^{(k)}_{s,j,l+m}\overline{{M}^{(k)}_{s',j',l'+m'}}C^{j,l+m}_{\tfrac 1 2, l, \tfrac 1 2, m}C^{j,l'+m'}_{\tfrac 1 2, l', \tfrac 1 2, m'}\right),
\end{align}
where
\begin{equation}
\hat{\rho}_{\rm mix}^{\otimes 2}(p)_{l,m,l',m'}=\left(D_{l,\tfrac 12}^{\tfrac 1 2}(\hat{U})D_{\tfrac 12,m}^{\tfrac 1 2}(\hat{U}^\dagger)+\delta_{l,\tfrac 1 2}\delta_{m,\tfrac 1 2}\right)\left(D_{l',\tfrac 12}^{\tfrac 1 2}(\hat{U})D_{\tfrac 12,m'}^{\tfrac 1 2}(\hat{U}^\dagger)+\delta_{l',\tfrac 1 2}\delta_{m',\tfrac 1 2}\right).
\end{equation}
After performing the integrations the result is
\begin{align}
&\int d {\mu}_U
	\bra{\psi}\Lambda[ \left((1-p)\ket{\psi}\bra{\psi}+p\ket{0}\bra{0}\right)^{\otimes 2}]\ket{\psi}=\nonumber\\
	&=\frac{p(1-p)}{3}\Sigma_k |M^{(k)}_{-\tfrac 1 2,0,0}|^2+\frac{p(1-p)}{6} \Sigma_k |M^{(k)}_{\tfrac 1 2,0,0}|^2+\frac{8(1-p)-5(1-p)^2}{12} \Sigma_k |M^{(k)}_{\tfrac 1 2,1,1}|^2+\nonumber \\ &+\frac{4(1-p)-3(1-p)^2}{12}\Sigma_k  |M^{(k)}_{-\tfrac 1 2,1,1}|^2+\frac{(1-p)^2}{4}\Sigma_k  |M^{(k)}_{-\tfrac 1 2,1,-1}|^2+\frac{(1-p)^2}{12}\Sigma_k  |M^{(k)}_{\tfrac 1 2,1,-1}|^2+\nonumber \\ &+
	\frac{(1-p)(1+p)}{6\sqrt{2}}\Sigma_k  |M^{(k)}_{-\tfrac 1 2,1,0}|^2+\frac{1-p}{6}\Sigma_k  |M^{(k)}_{\tfrac 1 2,1,0}|^2
	+\frac{p^2}{2}\nonumber \\ &+\frac{(1-p)^2}{6\sqrt{2}}\Sigma_k \Re [M^{(k)}_{\tfrac 1 2,1,0} \overline{{M}^{(k)}_{-\tfrac 1 2,1,-1}}]+\frac{(1-p)(1+p)}{6\sqrt{2}}\Sigma_k  \Re[M^{(k)}_{-\tfrac 1 2,1,0} \overline{{M}^{(k)}_{\tfrac 1 2,1,1}}].
\end{align}

Using the constraints (\ref{constraint2}) and the positivity and magnitude of the coefficients most of the optimal parameter choices can be found:
\begin{align}
\Sigma_k|M^{(k)}_{-\tfrac 1 2,0,0}|^2=1, \qquad\Sigma_k |M^{(k)}_{\tfrac 1 2,0,0}|^2=0&,\qquad \Sigma_k |M^{(k)}_{\tfrac 1 2,1,1}|^2=1, \nonumber\\ \Sigma_k  |M^{(k)}_{-\tfrac 1 2,1,1}|^2=0,\qquad\Sigma_k  |M^{(k)}_{-\tfrac 1 2,1,-1}|^2=1&,\qquad \Sigma_k  |M^{(k)}_{\tfrac 1 2,1,-1}|^2=0.\nonumber
\end{align}
Moreover, using the Cauchy-Schwartz inequality
\begin{align}
|\Re [M^{(k)}_{\tfrac 1 2,1,0} \overline{{M}^{(k)}_{-\tfrac 1 2,1,-1}}]|\leq \sqrt{\sum_k|M^{(k)}_{\tfrac 1 2,1,0}|^2}  \sqrt{\sum_k|M^{(k)}_{-\tfrac 1 2,1,-1}|^2}=\sqrt{\sum_k|M^{(k)}_{\tfrac 1 2,1,0}|^2} \nonumber \\
|\Re [M^{(k)}_{-\tfrac 1 2,1,0} \overline{{M}^{(k)}_{\tfrac 1 2,1,1}}]|\leq \sqrt{\sum_k|M^{(k)}_{-\tfrac 1 2,1,0}|^2}  \sqrt{\sum_k|M^{(k)}_{\tfrac 1 2,1,1}|^2}=\sqrt{1-\sum_k|M^{(k)}_{\tfrac 1 2,1,0}|^2}\;,
\end{align}
one is left with the maximisation of a function of the variable $t:=\sqrt{\sum_k|M^{(k)}_{\tfrac 1 2,1,0}|^2}$:
\begin{equation}
\frac{(1-p)(1+p)}{6\sqrt{2}}(1-t^2)+\frac{1-p}{6}t^2+\frac{(1-p)^2}{6\sqrt{2}}t+\frac{(1-p)(1+p)}{6\sqrt{2}}\sqrt{1-t^2}\;.
\end{equation}
The solution and the maximal value of the fidelity can be analytically determined, but they are quite cumbersome and we do not report them: instead we present the numerical plot 
 in Fig.~2 of the main text.

 \subsection{Upper Bound on Measurement and Prepare Protocols} 
We have already observed that in the limit of large $n_1$ and $n_2$, MP protocols allows for 
optimal average fidelity. But what happens for finite number of copies? 
To answer this question  we 
 introduce an upper bound on the average fidelity 
attainable with MP protocols.
Indeed, invoking once more the fact that for characterizing optimal performances one can restrict
the analysis to transformations which are  symmetric under the permutation of the first $n_1$ qubits. Using Eq.~(\ref{Fidelity 4}),
 the associated fidelity can be written as
\begin{eqnarray} 
F_{n_1,n_2}(\Lambda^{\text{MP}}):=\sum_{k=0}^{n_1}{n_1 \choose k} (1-p)^k p^{n_1-k} 
 \int d\mu_U \bra{\psi} \Lambda^{\text{MP}}(\ket{\psi}\bra{\psi}^{\otimes k}\otimes \hat{A}_{N-k})\ket{\psi},
\end{eqnarray}
where now $\Lambda^{\text{MP}}$ is the optimal MP channel. Then, we can get the following upper bound by using an optimal MP for each independent part of the whole state
\begin{eqnarray} 
F_{n_1,n_2}(\Lambda^{\text{MP}})&\leq&\sum_{k=0}^{n_1}{n_1 \choose k} (1-p)^k p^{n_1-k} \\ \nonumber
&\times& \int d\mu_U \bra{\psi} \Lambda^{\text{MP}}_k(\ket{\psi}\bra{\psi}^{\otimes k}\otimes \hat{A}_{N-k})\ket{\psi},
\end{eqnarray}
where $\Lambda^{\rm MP}_k$ is the optimal MP choice for $\ket{\psi}\bra{\psi}^{\otimes k}$. Using the known result for tomography of pure states \cite{Massar Pure} we can then derive the following inequality 
\begin{eqnarray} \label{UPPER} 
F_{n_1,n_2}(\Lambda^{\text{MP}})\leq&\sum^{n_1}_{k=0}{n_1 \choose k}\frac{k+1}{k+2} (1-p)^k p^{n_1-k},
\end{eqnarray}
where $\frac{k+1}{k+2}$ is the average fidelity in the optimal tomography of $k\geq 0$ copies of a pure state. Notice that the right-hand-side quantity
does not depend explicitly on $n_2$, and that for $n_1=2$ reduces to the function (19) of the main text which we reported in Fig.~\ref{figure2new}.
  \subsection{Performance of the Cirac, Ekert, Macchiavello (CEM)  protocol as a subtracting machine} 

 Here we show that a direct application of the method of Ref.~\cite{CIRAC} to solve our problem  for $n_1=2$ and arbitrary $n_2$
 leads to the same average fidelity as the DN strategy, being hence sub-optimal for our purposes. 

 The method presented in Ref.~\cite{CIRAC} does not assume the possibility of operating  on the noise signal, therefore the average fidelity one can achieve in this case does not depend on $n_2$. For case $n_1=2$, it consists of two steps  first performing an orthogonal measurement on the system  
 that discriminate the completely symmetric from the antisymmetric subspace of two qubits, and then tracing out on of the qubits. Adopting this procedure 
 from Eq. (4) of the main text we get
 \begin{eqnarray}
 F^{\rm CEM}_{n_1=2}&=&
 \!\!\!\int d {\mu}_U d {\mu}_V
 \bra{\psi}\Lambda^{\rm CEM}[ \hat{\rho}_{\rm mix}^{\otimes 2}(p)
 ]\ket{\psi}\;\\ \nonumber
 &=&\!\!\!\int d {\mu}_U d {\mu}_V
 \bra{\psi}\Lambda^{\rm CEM}[(1-p)^2\ket{\psi}\bra{\psi}^{\otimes 2} +p^2\ket{\phi}\bra{\phi}^{\otimes 2}+p(1-p)(\ket{\psi}\bra{\psi}\otimes\ket{\phi}\bra{\phi}+\ket{\phi}\bra{\phi}\otimes\ket{\psi}\bra{\psi})]\ket{\psi}\; .
 \end{eqnarray}
 Taking the integral on $\phi$ and using the fact that the method~\cite{CIRAC} is also covariant we can carry on the calculation
 \begin{eqnarray}
 F^{\rm CEM}_{n_1=2}&=&(1-p)^2+\frac{p^2}{2}+\!\!\!\int d {\mu}_U
 \bra{\psi}\Lambda^{\rm CEM}[p(1-p)(\ket{\psi}\bra{\psi}\otimes\frac{\hat{I}}{2}+\frac{\hat{I}}{2}\otimes\ket{\psi}\bra{\psi})]\ket{\psi}\;\\\nonumber
 &=&(1-p)^2+\frac{p^2}{2}+\bra{0}\Lambda^{\rm CEM}[p(1-p)(\ket{0}\bra{0}\otimes\frac{\hat{I}}{2}+\frac{\hat{I}}{2}\otimes\ket{0}\bra{0})]\ket{0} = 1-\frac{p}{2}\;.
 \end{eqnarray}
 where in the last inequality we use the fact that, as anticipated, $\Lambda^{\rm CEM}$ consists in 
 performing the measurements on the symmetric and antisymmetric subspace. We notice hence that $F^{\rm CEM}_{n_1=2}$ exactly coincides with the fidelity one would
 get by simply adopting the DN strategy, i.e. $F^{\rm CEM}_{n_1=2}= F^{\rm DN}_{n_1,n_2}$ which  is clearly not optimal in our case. 
  \end{widetext}


\begin{thebibliography}{1}

	\bibitem{WERNER} R. F. Werner, Springer Tracts in Modern Physics, Eprint arXive: 	arXiv:quant-ph/0101061.
	
	
	\bibitem{Wootters1982} W. K. Wootters and W. H. Zurek, Nature {\bf 299}, 802  (1982).
	
	
	\bibitem{dieks1982communication} D. Dieks, Phys. Lett. A {\bf 92}, 271 (1982).
	
	\bibitem{Noprogramming} M. A. Nielsen and I. L. Chuang, Phys. Rev. Lett. {\bf79}, 321 (1997).

	\bibitem{KumarPati2000} A. Kumar Pati and S. L. Braunstein, 
	Nature {\bf 404}, 164 (2000).
	
	
	\bibitem{PhysRevLett.100.090502} M. Piani, P. Horodecki, and R. Horodecki, Phys. Rev. Lett. {\bf 100}, 090502 (2008).
	
	\bibitem{QA2} U. Alvarez-Rodriguez, M. Sanz, L. Lamata, and E. Solano, Sci. Rep. {\bf 5}, 11983 (2015).
	
	\bibitem{PhysRevLett.116.110403} M. Oszmaniec, A. Grudka, M. Horodecki,
	and A. Wojcik, Phys. Rev. Lett. {\bf 116}, 110403 (2016).
	
	
	
	\bibitem{RevModPhys.77.1225} V. Scarani, S. Iblisdir, N. Gisin, and A. Acín, Rev. Mod. Phys. {\bf77} 1225 (2005).
	
		\bibitem{PhysRevA.96.052318} M. Doosti, F. Kianvash, and V. Karimipour, Phys. Rev. A {\bf 96}, 052318 (2017).
	
	\bibitem{PhysRevA.97.052330} S. Dogra, G. Thomas, S. Ghosh, and D. Suter, Phys. Rev. A {\bf 97}, 052330 (2018).
	
	
	\bibitem{Sepehr}
	P. Hayden, S. Nezami, S. Popescu, G. Salton,	arXiv:1709.04471 [quant-ph].
	
	
	

	
	\bibitem{ReviewFramesInfo} S. D. Bartlett, T. Rudolph, and R. W. Spekkens, Rev. Mod. Phys. \textbf{79}, 555 (2007).
	
	

	\bibitem{ResourceAsymmetry} G. Gour, and R. W. Spekkens, New J. Phys.  \textbf{10}, 033023 (2008).
	
	\bibitem{longevitycovariant}J. C. Boileau, L. Sheridan, M. Laforest, and S. D. Bartlett, Journal of Mathematical Physics {\bf49}, 032105 (2008).
	
	
	\bibitem{QM1} P. Wittek, {\it Quantum machine learning: what quantum computing means to data mining} (Academic Press, 2014).
	\bibitem{QM2}V. Dunjko and H. J. Briegel, Eprint arXiv:1709.02779. 
	\bibitem{QM3}  J. Biamonte, P. Wittek, N. Pancotti, P. Rebentrost, N. Wiebe and S. Lloyd, Nature {\bf 549}, 195-202 (2017).
	
	
	\bibitem{purificationprobabilistic}J. Fiur\'{a}\v{s}ek, New J. of Phys. {\bf 8}, 192 (2006).

	\bibitem{PhysRevLett.97.250503} G. Chiribella and G. M. D'Ariano, Phys. Rev. Lett. {\bf 97}, 250503, (2006).
	
	\bibitem{PhysRevLett.81.2598} D. Bruss, A. Ekert, and C. Macchiavello, Phys. Rev. Lett. {\bf 81}, 2598 (1998).
	
	\bibitem{chefles1998quantum} A. Chefles and S. M. Barnett, Journal of Physics A: Mathematical and General {\bf 31}(50), 10097, (1998).
		
		

	
	\bibitem{HOLEVOBOOK} A. S. Holevo, {\it Quantum Systems, Channels, Information:A Mathematical Introduction} (De Gruyter  2012).
	
	
	\bibitem{DEP1} M. A. Nielsen and I. L. Chuang, {\it Quantum Computation and Quantum Information} (Cambridge: Cambridge University Press, 2010).
	
	
	\bibitem{scarani3}
	V. Scarani et al., Phys. Rev. Lett. 88, 097905 (2002)
	
		\bibitem{CIRAC} 	
	J. I. Cirac, A. K. Ekert, and C. Macchiavello, Phys. Rev. Lett. {\bf 82}, 4344 (1999). 
	
	
	
	\bibitem{DEP2} C. King, 		
	IEEE Trans. Inf. Theory {\bf 49},  221-229,  (2003).
	
	\bibitem{QCOMP} T. D. Ladd , F. Jelezko, R. Laflamme, Y. Nakamura, C. Monroe, and  J. L. O'Brien,
	Nature {\bf 464}, 45 (2010).
	
	\bibitem{QCOM} N. Gisin and R. Thew,
	Nat. Phot. {\bf 1}, 165 (2007).
	
	\bibitem{IPPO} M. Ippoliti, L. Mazza, M. Rizzi, and V. Giovannetti, Phys. Rev. A {\bf 91}, 042322 (2015).
	
	\bibitem{OPTIMAL1} V. Bu\'{z}ek and M. Hillery, Phys. Rev. A {\bf 54}, 1844 (1996).
	
	\bibitem{OPTIMAL2} R. F. Werner, Phys. Rev. A {\bf 58}, 1827 (1998).
	
	\bibitem{OPTIMAL3} N. Gisin and S. Massar, Phys. Rev. Lett. {\bf 79}, 2153 (1997).
	
	\bibitem{OPTIMAL4} D. Bruss, A. Ekert, and C. Macchiavello, Phys. Rev. Lett. {\bf 81}, 2598 (1998).
	
	\bibitem{OPTIMAL5} V. Scarani, S. Iblisdir, N. Gisin, and A. Ac\'{i}n, Rev.  Mod. Phys. {\bf 77}, 1225 (2005).
	
	\bibitem{COVA} A. S. Holevo, Eprint 	arXiv:quant-ph/0212025.
	
	\bibitem{Massar Pure} S. Massar and S. Popescu, Phys. Rev. Lett. {\bf 74},	1259 (1995)
	\bibitem{FultonHarris} W. Fulton and J. Harris, Representation theory. A first course. Graduate Texts in Mathematics, Readings in Mathematics. 129. New York: Springer-Verlag (1991).
	
	\bibitem{SDP} L. Vandenberghe, and S. Boyd, SIAM Review Vol. 38, No. 1 (Mar., 1996), pp. 49-95 
	\bibitem{Mathematica} Wolfram Research, Inc., Mathematica, Version 11.3, Champaign, IL (2018).
	\bibitem{MATLAB}MATLAB and Statistics Toolbox Release 2018b, The MathWorks, Inc., Natick, Massachusetts, United States.
	\bibitem{CVX1} M. Grant and S. Boyd. CVX: Matlab software for disciplined convex programming, version 2.1. \url{http://cvxr.com/cvx}, December 2018.
	\bibitem{CVX2} M. Grant and S. Boyd. Graph implementations for nonsmooth convex programs, Recent Advances in Learning and Control (a tribute to M. Vidyasagar), V. Blondel, 					S. Boyd, and H. Kimura, editors, pages 95-110, Lecture Notes in Control and Information Sciences, Springer, 2008. \url{http://stanford.edu/~boyd/graph_dcp.html}. 
	\bibitem{Knapp}A. W. Knapp, Lie Groups Beyond an Introduction Birkhauser-Verlag, Basel, (1996).
	\bibitem{SUPMAT} Details of the calculations are available in the Supplemental Material at [URL will be inserted by publisher], where \cite{Knapp} is cited.
\end{thebibliography}
\end{document}